\documentclass[]{aa}
\usepackage{graphicx}
\usepackage[varg]{txfonts}
\usepackage{natbib}
\bibpunct{(}{)}{;}{a}{}{,} 
\usepackage[figuresright]{rotating}
\usepackage[a4paper,breaklinks]{hyperref}

\def\teff{${T}\rm_{eff}$ }
\def\kms{\,$\mathrm{km\, s^{-1}}$}

\newcommand{\loggf}{\ensuremath{\log\,gf}}
\newcommand{\logg}{\ensuremath{\log g}}
\newcommand{\cm}{cm$^{-2}\,$}

\newcommand{\kH}{$S_{\!\rm H}$}
\newcommand{\eu}[5]{\mbox{$#1\,^#2{\rm #3}^{#4}_{\rm #5}$}}
\newcommand{\Eexc}{$E_{\rm exc}$}
\newcommand{\eps}[1]{\log\varepsilon_{\rm #1}}
\newcommand{\mlp}{\ensuremath{\alpha_{\mathrm{MLT}}}}

\newcommand{\draftflag}{false}

\newcommand{\beq}{\begin{equation}}
\newcommand{\eeq}{\end{equation}}

\newcommand{\COBOLD}{{\sf CO$^5$BOLD}}

\newcommand{\cobold}{\COBOLD}

\newcommand{\xx}{\ensuremath{\mathrm{1D}_{\mathrm{LHD}}}}

\idline{10}{10}

\begin{document}

\title{A primordial star in the heart of the Lion\thanks{Based
on observations obtained at ESO Paranal Observatory,
GTO programme 086.D-0094 and programme 286.D-5045}}

\author{
E.~Caffau,\thanks{Gliese Fellow}\inst{1,2,3}
P.~Bonifacio,   \inst{2,3} 
P.~Fran\c cois, \inst{2,4} 
M.~Spite,       \inst{2} 
F.~Spite,       \inst{2} 
S.~Zaggia,     \inst{5} 
H.-G.~Ludwig,  \inst{1,2} 
M.~Steffen,    \inst{6} 
L.~Mashonkina, \inst{7}
L.~Monaco,     \inst{3}
L.~Sbordone,   \inst{1,2} 
P.~Molaro,     \inst{8} 
R.~Cayrel,     \inst{2} 
B.~Plez,       \inst{9} 
V.~Hill,       \inst{10}
F.~Hammer,     \inst{2} \and
S.~Randich     \inst{11}
}

\institute{ 
Zentrum f\"ur Astronomie der Universit\"at Heidelberg, Landessternwarte, 
K\"onigstuhl 12, 69117 Heidelberg, Germany
\and
GEPI, Observatoire de Paris, CNRS, Univ. Paris Diderot, Place
Jules Janssen, 92190
Meudon, France
\and
European Southern Observatory, Casilla 19001, Santiago, Chile
\and
UPJV, Universit\'e de Picardie Jules Verne, 33 Rue St Leu, F-80080 Amiens
\and
Istituto Nazionale di Astrofisica,
Osservatorio Astronomico di Padova, Vicolo dell'Osservatorio 5, 35122 Padova, Italy
\and
Leibniz-Institut f\"ur Astrophysik Potsdam, An der Sternwarte 16, 
D-14482 Potsdam, Germany
\and
Institute of Astronomy, Russian Academy of Sciences, RU-119017 Moscow, 
Russia 
\and
Istituto Nazionale di Astrofisica,
Osservatorio Astronomico di Trieste,  Via Tiepolo 11, 34143 Trieste, Italy
\and
Laboratoire Univers et Particules de Montpellier, LUPM, Universit\'e Montpellier 2, 
CNRS, 34095 Montpellier cedex 5, France
\and
Universit\'e de Nice Sophia Antipolis, CNRS,
Observatoire de la C\^ote d'Azur, Laboratoire Cassiop\'e e, B.P. 4229, 
06304 Nice Cedex 4, France
\and
Istituto Nazionale di Astrofisica,
Osservatorio Astrofisico di Arcetri, Largo E. Fermi 5, 50125 Firenze, Italy
}
\authorrunning{Caffau et al.}
\titlerunning{An extremely primordial star}
\offprints{E.~Caffau}
\date{Received ...; Accepted ...}

\abstract%
{
The discovery and chemical analysis of extremely
metal-poor stars permit a better understanding of
the star formation of the first generation of stars and of the Universe emerging
from the Big Bang.
}
{ 
We report the study of a primordial star situated in the centre of the
constellation Leo (SDSS\,J102915+172027).
}
{
The star, selected from the low resolution-spectrum of the Sloan Digital Sky Survey,
was observed at intermediate (with X-Shooter at VLT) and at high spectral 
resolution (with UVES at VLT).
The stellar parameters were derived from the photometry. The standard
spectroscopic analysis based on 1D ATLAS models
was completed by applying 3D and non-LTE corrections.
}
{ 
An iron abundance of [Fe/H]=--4.89 makes SDSS\,J102915+172927 one of the lowest
[Fe/H] stars known.
However, the absence of measurable C and N enhancements indicates that it
has the lowest metallicity, ${\rm Z}\le 7.40\times 10^{-7}$ 
(metal-mass fraction), ever detected.
No oxygen measurement was possible.
}
{
The discovery of SDSS\,J102915+172927 highlights that
low-mass star formation occurred at metallicities lower than
previously assumed.
Even lower metallicity stars may yet be discovered,
with a chemical composition closer to the composition of the primordial
gas and of the first supernovae.
}

\maketitle

\section{Introduction}

Very metal-poor stars are the relic of the primordial Universe.
They are formed from material whose composition is close to that
of the Universe emerging from the Big Bang, only with traces
of elements heavier than lithium, 
which were processed by the first generations of stars. 
A low-mass star (${\rm M<0.8 M_\odot}$) formed just after the
Big Bang would still shine today on the main sequence, 
and could be easily recognised from its observed spectrum
because only lines of hydrogen and the Li doublet at 670.7\,nm
would be visible.
Such stars have not been detected, and some models of star
formation state that it is impossible that such star might form.
Indeed, it has been speculated that the first generation of stars
consisted of massive stars (${\rm M>80 M_\odot}$ \citealt{ostriker96})
that evolved rapidly and
synthetised metals that were ejected into space for the following generations
of stars through a supernova type II explosion.
This scenario is supported by the fact that no zero-metal star has been 
observed so far. But absence of evidence is not  evidence of absence.
Such stars, if they exist, could be very rare and faint.
Some theories of star formation predict that the stars that formed from the
ejecta of the first few generations of primordial massive stars needed
a chemical pattern different from the one well-known in metal-poor stars
that, apart from an enhancement in $\alpha$-elements, have a scaled solar
composition \citep{bonifacio09}.
According to \citet{bromm03} and \citet{schneider03},
a minimum metallicity is necessary
to permit the cooling of the collapsing cloud for small-mass stars to form.
In particular, according to \citet{bromm03},
extremely iron-poor stars  can form 
at sub-solar masses when enhanced in C and O.
The discovery of some hyper-iron-poor and very
carbon- and nitrogen-enhanced stars in the past decade \citep{christlieb02,frebel05,norris07} 
supported the theory of \citet{bromm03}.
It is interesting to note that
the cooling by dust \citep{omukai08,schneider11} 
may imply a considerably lower critical metallicity,
with no specific requirement of chemical composition. 
Another fundamental ingredient of star formation
besides cooling is fragmentation. If a large-mass cloud
can fragment into smaller pieces, then  these
pieces can form low-mass stars even in the absence of metals, 
the main cooling agent being molecular hydrogen \citep{nakamura}.
Recently, \citet{clark11} have shown that the first generation
of stars are indeed likely to form as binary or multiple systems
resulting from the fragmentation of an initial cloud of large mass. 
Under these conditions, \citet{greif11} derived an initial mass
function (IMF) that is essentially flat between 0.1 and 10 solar masses.
This IMF is top-heavy, in the sense that most of the mass is concentrated
in a few massive stars, and yet it allows for the formation of low-mass
stars. In this latter scenario there is no critical metallicity.

We present here the analysis of SDSS\,J102915+172927,
an extremely metal-poor star, that shows no strong enhancement
in carbon and nitrogen, and has a global metallicity more than four
orders of magnitude lower than the solar one:
${\rm Z}\le 7.40\times 10^{-7}$, 
to be compared to ${\rm Z}_\odot =1.53\times 10^{-2}$ \citep{abbosun},
giving a metallicity in respect with solar one of ${\rm Z}=5~10^{-5}{\rm Z}_\odot$.
If CNO abundances are non-enhanced,
${\rm Z}= 7.15\times 10^{-7}$.
With respect to the results presented earlier by our group 
\citep{nature}, the present analysis uses of a custom
\cobold\ hydrodynamical model and NLTE computations for Mg, Si, Ca, Fe,
and Sr.

\section{Observations and data reduction}

For SDSS\,J102915+172927 we availed ourselves of X-Shooter \citep{dodorico} and UVES
\citep{dekker00} observations.
The star was observed during the X-Shooter French-Italian 
Guaranteed Time of Observation (GTO) on 10 February 2011, 
programme 086.D-0094, P.I. P. Bonifacio.
The spectral range covered is 330-2400\,nm, split into three spectral arms (UVB, VIS, and NIR).
For the integral field unit (IFU), which re-images an input field 
of 4"x 1.8" into a pseudo slit of 12"x 0.6" and a $1\times 2$ binning, staring mode
was used.
Using the IFU as a slicer, the resolving power was 
R=7900 in the UVB arm  and R=12600 in the VIS arm.
The spectra were reduced using the X-Shooter pipeline \citep{goldoni}.
The details on X-Shooter observations can be found in \citet{caffau11}.

On the basis of the highly promising analysis of the X-Shooter spectrum
we applied for ESO Director Discretionary Time (DDT) 
to observe the star with UVES@VLT,
and the observing blocks (OB) were observed on March-April 2011, 
within the programme 286.D-5045, P.I. P. Bonifacio.
Five OBs were executed in DIC \#1, standard 390nm + 580nm setting, 
two in DIC \#2, standard 473nm + 760nm setting.
All exposures had a length of 3000s, slit width was set to 1.4'', 
which led to an expected resolution of $\sim$38\,000.
However, during most exposures the seeing was better than 1.4'', 
thus leading to slightly better resolution in the final spectra.
The signal-to-noise ratio per pixel at 650\,nm was between 30 and 40 for all spectra.

\begin{table}
\caption{\label{star}
Characteristics of the star SDSS\,J102915+172927.
}
\begin{center}
\begin{tabular}{ll}
\hline\hline\noalign{\smallskip}
Star                           & SDSS\,J102915+172927\\  
RA (J2000.0)                   & $10h\, 29m\, 15.15s$\\
Dec (J2000.0)                  & $+17^\circ\, 29'\, 28''$\\
$l$                            & $221.785^\circ$\\
$b$                            & $+55.863^\circ$\\
Epoch of SDSS photometry       & 2005.929 \\
u, SDSS                        & $17.736\pm 0.011$\\
g, SDSS                        & $16.922\pm 0.004$\\
r, SDSS                        & $16.542\pm 0.004$\\
i, SDSS                        & $16.388\pm 0.004$\\
z, SDSS                        & $16.330\pm 0.008$\\
J, 2MASS                       & $15.513\pm 0.052$\\
H, 2MASS                       & $15.126\pm 0.077$\\
K, 2MASS                       & $15.146\pm 0.116$\\
A$_V$                          & 0.084\\
${\rm V_{\rm rad}}$, [\kms]    & $-34.5\pm 1.0$\\
${\rm T_{\rm eff}}$, [K]       & $5811\pm 150$\\
\logg                          & $4.0\pm 0.5$\\
$\xi$  [\kms]                  & 1.5\\
Metallicity, SDSS              & $-3.73$\\
$\left[{\rm Fe/H}\right]$      & $-4.89\pm 0.10$\\
$\left[\alpha /{\rm H}\right]$ & $0.23\pm 0.26$\\[4pt]
\hline\hline
\end{tabular}
\end{center}
\end{table}

\section{Model atmospheres}

The analysis was mainly performed with 1D plane-parallel hydrostatic model 
atmospheres, computed with ATLAS\,9 \citep{kurucz93,kurucz05} in its Linux 
version \citep{sbordone05}. The line opacity was treated through opacity 
distribution functions (ODFs), and we used the ODFs computed
by \citet{castelli03}  for a scaled solar metallicity corresponding
to [Fe/H]=-4.5. The ATLAS\,9 model was computed with the stellar parameters
derived from photometry, as given in Tab.\,\ref{star}.
The 3D abundance corrections were derived with a 3D-\cobold\ model
expressly computed for this star and the related \xx\ model,
computed with the LHD code \citep[for details concerning the 3D corrections 
see][]{abbosun}.
To derive the 3D corrections on the chemical abundances we compared the
abundance derived by using the 3D-\cobold\ model with the abundance derived
with 1D plane-parallel model, the \xx\ model. The \xx\ model employs the same
micro-physics and radiative transfer scheme as the 3D-\cobold\ model,
so that any difference in the abundance determination is due to the
3D versus 1D treatment of convection.
The 3D model used in the following was computed for stellar parameters 
\teff = $5850$\,K, \logg = $4.0$, [Fe/H] = $-4.0$. The computational
domain is a cartesian box of size $25.8\times 25.8\times 12.5$\,Mm$^3$, 
resolved by $200\times 200\times 200$ cells. 
The frequency dependence of the total radiative opacity 
was approximately taken into account by adopting the opacity binning method 
(see e.g. \citealt{freytag12}), using a total of $11$ frequency 
groups. The detailed monochromatic (continuous plus line) opacities needed to 
evaluate the average opacity for each frequency group as a function of $P$ 
and $T$ were derived from the MARCS stellar atmosphere package \citep{gustaffson03,gustaffson08}
for metallicity [Fe/H]=$-4$, [$\alpha$/H]=0.4 
(B. Plez, priv. comm.).

\begin{figure}
\begin{center}
\resizebox{\hsize}{!}{\includegraphics[draft = \draftflag,clip=true]{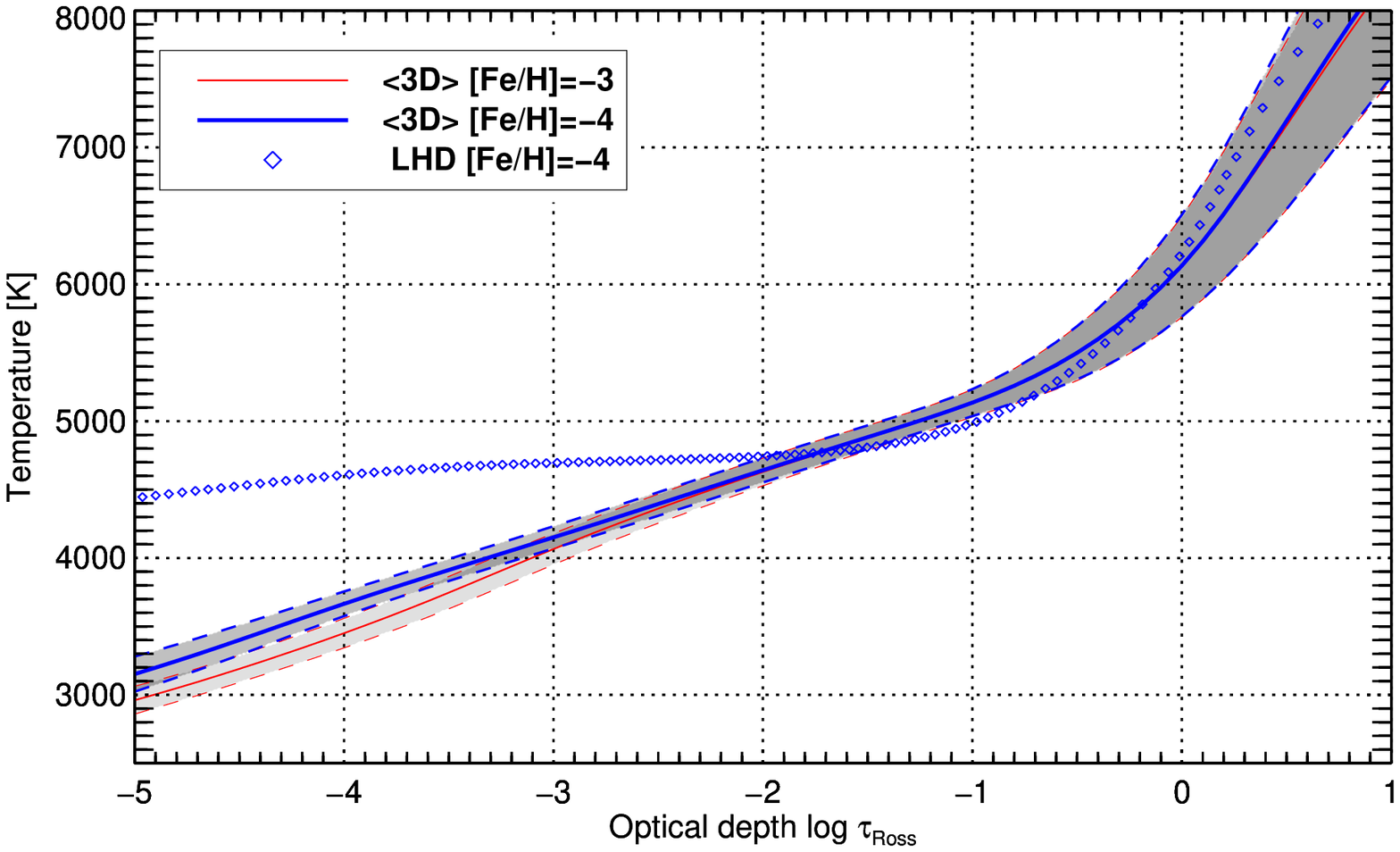}}
\end{center}
\caption[]{Mean temperature structure as a function of Rosseland optical
depth of the \cobold\ model used in this work (\teff=$5850$~K, \logg=$4.0$, 
[Fe/H]=$-4.0$; thick [blue] line), and of the 3D model used in \cite{nature} 
($5850/4.0/-3.0$; thin [red] line). For each model, the mean temperature is 
obtained by averaging the temperature as $\sqrt[4]{\langle{T^4}\rangle}$ 
over iso-$\tau_{\rm Ross}$ surfaces of $20$ 
selected snapshots. Shaded areas bounded by dashed lines indicate the 
$\pm 1\sigma$ range of deviations. Small open symbols outline the temperature
stratification of the \xx\ model with [Fe/H]=-4.}
\label{ttau}
\end{figure}

The present 3D model differs from the one previously used for the 
calculation of 3D abundance corrections in \citet{nature} in several
aspects: (i) its metallicity is lower by a factor of 10
([Fe/H]=$-4$ instead of $-3$), (ii) it uses a better frequency resolution 
($11$ instead of $6$ frequency groups), (iii) it has a higher spatial
resolution ($200\times 200\times 200$ instead of $140\times 140\times 150$
grid cells). Nevertheless, the mean temperature stratifications are quite
similar (see Fig.\,\ref{ttau}), except for the upper photosphere 
($\log \tau_{\rm Ross} < -3$) where -- contrary to expectations -- the
lower metallicity model shows somewhat higher temperatures. 
Both 3D models agree in the amplitude of the photospheric temperature 
fluctuations, which is fairly low, presumably because of H$_2$ molecule formation 
in the higher photosphere.
The most striking difference between the 3D and the 1D models
is the temperature in the outer photosphere, which in both 3D models
is significantly cooler  than their \xx\ counterparts. 

The total 3D LTE abundance corrections (A(3D) - A(\xx))
were computed for all atomic lines measured
in the observed spectra, assuming \mlp=0.5, and a microturbulence
parameter $\xi=1.5$\kms\ for the \xx\ models. 
For the molecular bands, G-band and NH-band, 
representative lines were considered.
Obviously, the physical reason for the large negative 3D corrections
for \ion{Fe}{i} and \ion{Ni}{i} is the large 3D-1D temperature difference
in the upper photosphere.

\section{Analysis}

The reference solar abundances we used were taken from \citet{abbosun}
for C, N, O, and Fe, and from \citet{lodders09} for the other elements.
The atomic data for the lines were taken from the large programme ``First Stars'' 
lead by Roger Cayrel \citep{bonifacio09,cayrel04,francois07}.

\subsection{Stellar parameters}

We derived the effective temperature
of 5811\,K from the $\left(g-z\right)$ colour \citep{ludwig08}.
We also fitted the H$\alpha$ wings. By using a grid of synthetic spectra
computed from an ATLAS model and a modified version of the {\tt BALMER} 
code\footnote{The original version is available on-line at \url{http://kurucz.harvard.edu/}}
that uses the theory of \citet{barklem00,barklem00b}
for the self-broadening and the profiles of \citet{stehle99} for
Stark broadening, we derived an effective temperature within few K.
However, when using a grid of 3D profile the temperature is about
100\,K hotter.
We assumed an uncertainty of 150\,K in the effective temperature.
This value is derived by linearly adding the uncertainties in the SDSS $(g-z)$ colour
and an estimate on the error on  the zero point of the calibration.
We took 75\,K as the uncertainty due to the error in the colour on the 
derived \teff, which corresponds to three times $\sigma$=25\,K.
Comparing the \teff\ derived from the $(g-z)$ colour 
of the four calibrator stars of \citet{fukugita96} and
the IRFM temperatures of 
\citet{alonso}, we find an average difference of about 70\,K.
A similar comparison with the IRFM temperatures of \citet{GB} provides
a difference of the order of 10\,K. Conservatively, we adopted 
the difference with the \citet{alonso} temperatures as
an estimate of the zero point error of the calibration.

UVES spectra were able to give us better insight than the X-Shooter data
in the gravity determination. The
Ca ionisation equilibrium, when computed in LTE, supports  
the adopted gravity (\logg = 4.0).
By comparing the $(u-g)$ colour with the theoretical
colours computed from the grid of ATLAS 9 models
of \citet{castelli03}\footnote{wwwuser.oats.inaf.it/castelli}, we derived \logg = 3.9.
The use of the $u-g$ colour for gravity determination
implies a fairly large uncertainty, both for the sensitivity to photometric errors
(an error of  0.05 mag in the colour translates into an error of 0.3\, dex in \logg)
and for the uncertainty in the zero point. 
A comparison of the $u-g$ derived gravities with the Hipparcos parallax
based gravities of \citet{GB} for the
four SDSS photometry calibrators \citep{fukugita96} suggests that the
$u-g$ colour provides gravities that are systematically lower by 0.5\,dex.
We cannot exclude that the star has indeed a higher gravity of about 4.5. 
The non-LTE \ion{Ca}{i}/\ion{Ca}{ii} equilibrium would require an even higher gravity
of about \logg = 4,8, which is difficult to accept considering the gravities 
derived from isochrones, which are compatible with the effective temperature of the star.
Nevertheless, all this evidence allows us to robustly exclude that the star is
either a sub-giant or a horizontal-branch star.

No \ion{Fe}{ii} line is detectable in the spectrum, therefore the iron ionisation
equilibrium cannot be adopted for gravity determination. But lines of both
ionisation states of Ca are present in the UVES spectrum.
The 1D-LTE abundance derived from the three lines of the IR \ion{Ca}{ii} triplet
differs by 0.12\,dex from the abundance derived from the \ion{Ca}{i}
line at 422.6\,nm in the 1D-LTE analysis. This latter line is insensitive to gravity,
while the IR \ion{Ca}{ii} lines depends on the choice of gravity. 
This good agreement between the Ca lines could be an indication that the gravity is correct,
but it is broken when we apply non-LTE corrections.
In this case the difference becomes 0.38\,dex. 
The use of ${\rm S}_{\rm H} = 1$ instead of the adopted value
of 0.1 (see section\,\ref{sec:nlte})  
cannot remove the difference between the two ionisation stages, 
although it does decrease (0.23\,dex).
The higher Ca abundance from the
neutral \ion{Ca}{i} line would require higher gravity.
When considering a gravity of \logg =4.5,
the Ca abundance from the \ion{Ca}{i} line remains
mostly the same (within 0.03\,dex) both in 1D-LTE and 1D-NLTE, while
the abundance of Ca derived from the IR-triplet would increase by 0.12 and 0.17\,dex
in 1D-LTE and 1D-NLTE, respectively.
The agreement would definitely improve, but the increase in gravity would not be
enough to result in an agreement of the Ca abundance derived from the
two ionisation states of Ca, and we cannot support a higher value of gravity
in this star.
To determine the gravity we also fitted the
wings of high-order Balmer lines (H$\epsilon$ and higher) and found that
the gravity should be in the range $4.3\le$\logg $\le 3.6$
when the temperature of the synthetic spectra is fixed at 5811\,K.
By increasing the temperature by about 200\,K, the gravity increases by 0.5\,dex.
From all these results we decided to keep \logg =4.0 as our value for the gravity,
but the star could be a main-sequence star with \logg =4.5 or a turn off-star
with \logg $>$ 3.5. We are confident that the star is no sub-giant.
We cannot exclude that the problem with the Ca abundance is related to
the effective temperature.
An increase of $\pm 200$\,K in effective temperature would change the calcium abundance
derived from the 422.6\,nm line by about $\pm 0.2$\,dex, 
and the abundance derived from the \ion{Ca}{ii} lines
of the triplet would change by about $\pm 0.1$\,dex.

We derived the micro-turbulence of 1.5\kms\ 
by using the formula in \citet{edvardsson93}.
The same value of the micro-turbulence was applied to the \xx\ model
used for computing the 3D corrections.
The 3D corrections are sensitive to the choice of the micro-turbulence,
because the abundance from the \xx\ model depends on it in the same 
way as the abundance derived from the ATLAS model.
The 1D abundance corrected by 
the 3D corrections is therefore largely  
insensitive to the choice of micro-turbulence.

The radial velocity of the star from the UVES spectrum is $-34.5$\kms,
which agrees well with the radial velocity derived from X-Shooter
and SDSS spectra.

\subsection{Variability}

We investigated possible long term-photometric variabilities of SDSS\,J102915+172927
and, given the material at our disposal, we can exclude this at a level
of $\Delta r\le0.006$.
A detailed description can be found in the appendix.

\subsection{Distance (isochrone fitting)}

We fitted the eight-band photometry (SDSS+2MASS) as reported in Table~\ref{star} of
SDSS\,J102915+172927 with different sets of low-metallicity isochrones.
There are no published isochrones that match the low metallicity of 
this type of object. We collected a sample of the most recent low-metallicity isochrones
and added two special cases: \citet{marigo03}  
who presented isochrones for zero-metallicity low-mass stars but limited to $0.70$M$_\odot$; 
and an unpublished set of FRANEC isochrones,
computed according to the prescriptions
of \citet{SCL}, kindly provided by Chieffi and Limongi, 
with [Fe/H]$=-6.0$. The sets of all theoretical isochrones used are 
listed in Table~\ref{isonane}.

\begin{figure}
\begin{center}
\resizebox{\hsize}{!}{\includegraphics[angle=270,draft = \draftflag,clip=true]%
{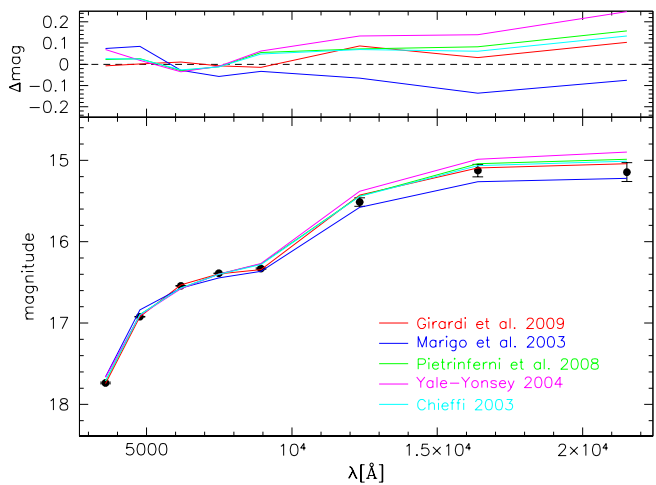}}
\end{center}
\caption[]{Best solutions of the fitted isochrones to the eight-band SDSS-2MASS 
photometry of SDSS\,J102915+172927. In the upper panel we depict the solution residuals.}
\label{iso}
\end{figure}

\begin{table*}
\caption{\label{isonane}
Parameters of the best-fitted ($\chi^2$) isochrones for main-sequence stars.}
\begin{tabular}{lcccllllcc}
\hline\noalign{\smallskip}
Author & [Fe/H] & Y & $\left[\alpha /{\rm Fe}\right]$ & Age & Mass & \logg & ${\rm T_{\rm eff}}$ & (m-M)$_0$ & $\chi^2$\\
                            &      &         &[dex]  & [Gyr]& [M$_\odot$]& & [K] & [mag] &\\[4pt]
\hline\noalign{\smallskip}
Marigo et al.        & Zero   & 0.249 & $-$   & 12.6 & 0.70 & 4.6 & 6108.0 & 11.09 & 0.070\\
Chieffi \& Limongi   &$-6.00$ & 0.232 & $-$   & 12.0 & 0.68 & 4.7 & 5775.3 & 10.68 & 0.029\\
Pietrinferni et al   &$-3.26$ & 0.245 & +0.40 & 12.5 & 0.64 & 4.6 & 5751.5 & 10.51 & 0.033\\
Yale--Yonsey         &$-3.76$ & 0.230 & +0.60 & 13.0 & 0.62 & 4.7 & 5592.4 & 10.29 & 0.038\\
Girardi et al.       &$-2.45$ & 0.245 & +0.00 & 13.0 & 0.67 & 4.6 & 5873.5 & 10.76 & 0.011\\
\noalign{\smallskip}\hline\noalign{\smallskip}
\end{tabular}
\end{table*}

The fitting was made using all photometric measurements from
$u$ to $K$ weighting on the photometric precision. This is particularly important
since the infrared 2MASS colours of this star are at the lower limits of detection
and should be treated with care. For each stellar mass of the isochrone we varied
only the distance modulus to calculate $\chi ^2$. We then chose the
best model on the basis of the lower $\chi ^2$. Considering that the gravity from
spectroscopy is compatible with a dwarf star, we assumed that the star is on the
main sequence. The age range assumed is $12.0$ or $13.0$~Gyr depending on the availability
of the isochrones. We checked the goodness of the solutions also along the sub-giant
branch, but we always found that the solutions were worse than the dwarf solutions.
All isochrones used are on the VEGA
Johnson-Cousins system, except for the Girardi  isochrones
\citep{G02,G04,G05}, which can be obtained for a variety of
different photometric systems and for which we chose SDSS+2MASS directly.
To obtain all magnitudes on this system, we properly transformed the
isochrones in the observed system using the colour 
transformations provided by the 
SDSS\footnote{http://www.sdss3.org/dr8/algorithms/sdssUBVRITransform.php}.
This of course has some effect on the given precision of
the solutions. In the process of fitting the photometry, each isochrone was
reddened with the observed value of extinction given by the Schlegel maps
(see Table~\ref{star}) using the proper extinction coefficients for the SDSS/2MASS
passbands. The output of the whole procedure is the value of the distance modulus
and the stellar mass of the isochrone.

The calculated best solution together with the residuals are plotted in Fig.\,\ref{iso} where
we show the five isochrones used in this study. In Table~\ref{isonane} we list
the parameters for each of the best solutions together with the $\chi ^2$ value.
In the figure the solutions overimposed 
on the photometric measurements are shown in the lower panel,
while the residuals of the fitting are plotted in the upper panel.
While for four isochrones we obtained the absolute best solutions for the distance
and mass, for the \citet{marigo03} data 
our search stopped at the minimum stellar mass available, $0.70$~M$_\odot$, but 
clearly the fitting trend required a lower stellar mass of the order of 
$\simeq0.60$~M$_\odot$. Future versions of these isochrones (Bressan et al., in preparation)
will allow a more precise constraint of the evolutionary status of this star.

Finally, it is not surprising to find that the best isochrone is that of Girardi et al.
because it was ``coloured'' directly in the SDSS/2MASS systems. 
Our preference is the Chieffi\& Limongi isochrone because it gives the lowest 
$\chi^2$, the best match with the spectroscopic metallicity, and with the
${\rm T_{\rm eff}}$ of the star (a difference of only $\simeq35^\circ$~K). 
To summarise, the assumed distance modulus is $\left({\rm m-M}\right)_0=10.68\pm0.15$\,mag 
or $1.37\pm 0.20$\,kpc. 

\subsection{Kinematics}

Kinematical data are fundamental for understanding the nature of
SDSS\,J102915+172927. Radial velocity and proper motion
measurement allowed us  to compute a Galactic orbit.

\subsubsection{Radial velocity}

The radial velocity of the star as measured from the summed spectrum 
of all UVES data is $-34.5\pm 1.0$~\kms, which agrees well
with the $-31.0\pm 10.0$~\kms measured from the X-Shooter
spectrum. We found no significant variations of the radial velocity.
The difference of UVES and X-Shooter is well within the given error estimates.
Given the low value measured, we performed an analysis of the compatibility 
of the radial velocity with the different stellar populations of the Galaxy.
We used the Besan\c con Galactic model \citep{Robin} to
simulate a  field of 1~square degree
surrounding the star. In the simulated field we selected a sample of stars 
with similar magnitude and colours 
around SDSS\,J102915+172927: we used a box  
$0.1$~mag wide in both $r$ and $(r-i)$. The observed radial 
velocity lies within 1.2$\sigma$ from the expected mean radial velocity 
of the field stellar population, thin and thick disc and halo, 
which has an average of 
$+37.4$~\kms\ and a $\sigma=58.2$~\kms. 
A ``normal'' radial velocity excludes the possibility 
that this star is 
a member of recently accreted streams or other substructures 
present in the halo of the Galaxy.

\begin{figure}
\begin{center}
\resizebox{\hsize}{!}{\includegraphics[angle=270,draft = \draftflag,clip=true]%
{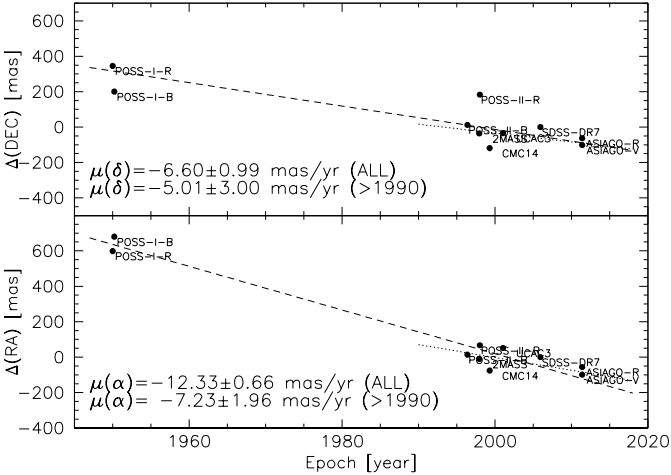}}
\end{center}
\caption[]{Measurement of the proper motion in the two coordinates, RA in the bottom panel, 
DEC in the upper panel. The zero points for the offsets are given by the 
position in the SDSS-DR7 catalogue. The results of the two different fits are shown: 
for all positions (long dashed line) and for the positions since 1990 only (dotted line).}
\label{propmot_102915}
\end{figure}
 
\subsubsection{Proper motion}

Although recent surveys and old photographic catalogues have allowed us to build 
all-sky catalogues of proper motions for several millions of stars, subtle 
systematic errors are always lurking in the data, therefore we
recommend caution.
 
In the present case the recent PPMXL Catalogue \citep{roeser10} provides  
a proper motion measurement for this star: 
$\mu _\alpha{\rm cos}(\delta) = -12.8\pm3.9$~mas/yr and 
$\mu _{\delta} = -6.7\pm 3.9$~mas/yr. The values are based on six different measurements
comprising POSS~I and II, 2MASS, and SDSS. This is not the only 
available measurement since also SDSS-DR7 \citet{Munn04}
provides a measurement for the proper motion:  
$\mu _\alpha{\rm cos}(\delta) = -19\pm2$~mas/yr and $\mu\delta=-1\pm2$~mas/yr. 
There also seems to be a negative value for the right ascension here,
but almost no motion in declination. However, the latter measurements were 
discovered to have systematic errors 
\citep{Munn08} 
that were corrected, but leave some doubt about the derived proper motions.
We have to note that the faintness of
the object makes it intrinsically difficult to 
measure in some of the surveys (like 
in 2MASS and UCAC3) because it is near the magnitude limit of the detectors. 

To check the validity of the above values of proper motions, we independently
measured the positions of the star in images of different epochs. 
We then combined these positions  with 
the star's positions in different recent 
astrometric catalogues based on CCD detectors.  
The proper motion strongly depends on the 
first epoch position given by the POSS-I
plates taken in the 1950s. Since then there is a long gap until the 1980s. 
To improve the first epoch position, we recalculated the astrometry
of the available photographic plates, POSS-I and POSS-II with the help of the 
SDSS photometry in the field. We recovered $1\times1$~deg$^2$ images for both 
POSS-I/II  around our target star in the 
blue and red from the MAST database 
at STScI. We then obtained a new astrometric solution
for each POSS image using the SCAMP 
programme\footnote{\label{bertin}E. Bertin, see for 
details http://www.astromatic.net/}.
The extracted sources were compared with an
SDSS-DR7 reference catalogue built by selecting only
objects classified as compact galaxies 
with magnitude $r<21.00$. A total of 2532 galaxies were
used as reference out of a total of 20532 objects present in the field.
With this sample of galaxies we minimised the 
systematic effects present in independent astrometries. We then ``swarped'' 
each image with 
the SWARP programme
(see footnote \ref{bertin}) and measured the positions with SEXTRACTOR
(parameters X/Y-WIN\_IMAGE and given errors).
We found no significant 
presence of systematic trends either with position 
in the plate, magnitude, or type of object. 
We verified that the used reference galaxies
defined a common zero point of the two astrometries (POSS and SDSS-DR7).

A similar procedure has been used also on the recent R- and V-band images taken
at the Asiago Cima Ekar Schmidt Telescope 
already described above. We used the same
reference list of galaxies to perform the 
astrometrisation of the field, again using the 
SCAMP/SWARP/SEXTRACTOR sequence of programmes. 
In this case we started for each band
from the four images and ``swarped'' 
the resulting final positions 
with sextractor to a single image.

To the 2~POSS-I, the 2~POSS-II, and 
two~Asiago positions we added 2MASS, UCAC3, CMC14
and, of course, SDSS-DR7 to compute the final proper motions that are 
shown in Figure~\ref{propmot_102915}. 
Clearly, the entire ``signal'' of the proper
motion is indeed given by the first epoch positions of the 2 POSS-I plates. 
Fitting linear regression gives a solution of 
$\mu_\alpha \cos(\delta) = -12.33\pm0.66$~mas/yr and 
$\mu_\delta=-6.60\pm0.99$~mas/yr, 
which is quite similar to the PPMXL solution.
If we drop the POSS-I and use 
only the measurements taken after 1990.0, we find
a less significant solution 
$\mu_\alpha \cos(\delta) = -7.23\pm1.96$~mas/yr and 
$\mu_\delta=-5.01\pm3.00$~mas/yr.

\begin{figure}
\begin{center}
\resizebox{\hsize}{!}{\includegraphics[draft = \draftflag,clip=true]%
{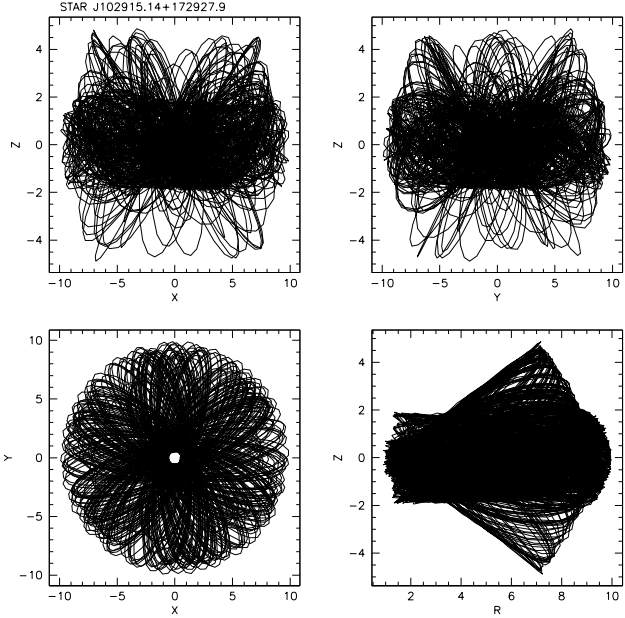}}
\end{center}
\caption[]{Galactic orbit for star J102915.14+172927.9 obtained using the
PPMXL proper motion. The top-left 
panel shows the X/Z plane; top-right shows the Y/Z plane; bottom-left 
shows the X/Y plane; bottom-right shows the meridional plane R/Z.}
\label{orbit_102915}
\end{figure}

\subsubsection{Galactic orbit}

Combining the positional and kinematic information, we
calculated the Galactic orbit of the star by performing several
tests varying the input parameters within the errors. 
The code used (Gratton et al. 2003) is based on the Allen \& Santillan 
(1991) prescriptions for the Galactic potential components with updated 
parameters for the constants describing the potential.
Although the low height on the Galactic plane ($Z\sim1.0$~kpc) may suggest a thick disk
orbit, this can be safely ruled out. The orbit solution indicates that
the star belongs to the halo with the maximum height above the
galactic plane ${\rm Z_{\rm max}}=4.8\pm 0.4$\,kpc, the orbital
apocenter at ${\rm R}_{\rm max}=9.6\pm 0.6$\,kpc, and is plunging
towards the Galactic centre, with orbital pericenter ${\rm R}_{\rm
min}=0.9\pm 0.1$\,kpc. See Fig.\,\ref{orbit_102915}.
Adopting the proper motion values obtained in the previous section from
the positions after 1990.0, we obtain a similar orbit with a more extreme
orbital pericenter ${\rm R}_{\rm min}=0.4\pm 0.1$\,kpc. 
An even more extreme value of $0.2\pm0.1$~kpc is obtained when
we adopt a null value of the proper motion.

\subsection{Abundances}

Very few lines are measurable in the X-Shooter spectrum. The \ion{Mg}{i}-b
triplet is not visible. Of the IR \ion{Ca}{ii} triplet lines,
only the one at 854.2\,nm is clearly visible, but is contaminated 
by a feature produced by the sky subtraction.
Some \ion{Fe}{i} lines can be guessed, not really measured.
The only clearly detectable line is the \ion{Ca}{ii}-K line at 393.3\,nm.
Its EW of 49.2\,pm is consistent with an abundance of [Ca/H]=--3.9.
But the measured radial velocity is --30\kms, comparable to the 
X-Shooter UBV arm resolution of 7\,900, which suggests that the line
is contaminated by the component from the interstellar medium (ISM).
From the X-Shooter spectrum,
we can deduce that this spectrum belongs to an extremely metal-poor star and
put an upper limit on the metallicity of about --4.0
with respect to the solar metallicity.

\begin{figure}
\begin{center}
\resizebox{\hsize}{!}{\includegraphics[draft = \draftflag,clip=true]%
{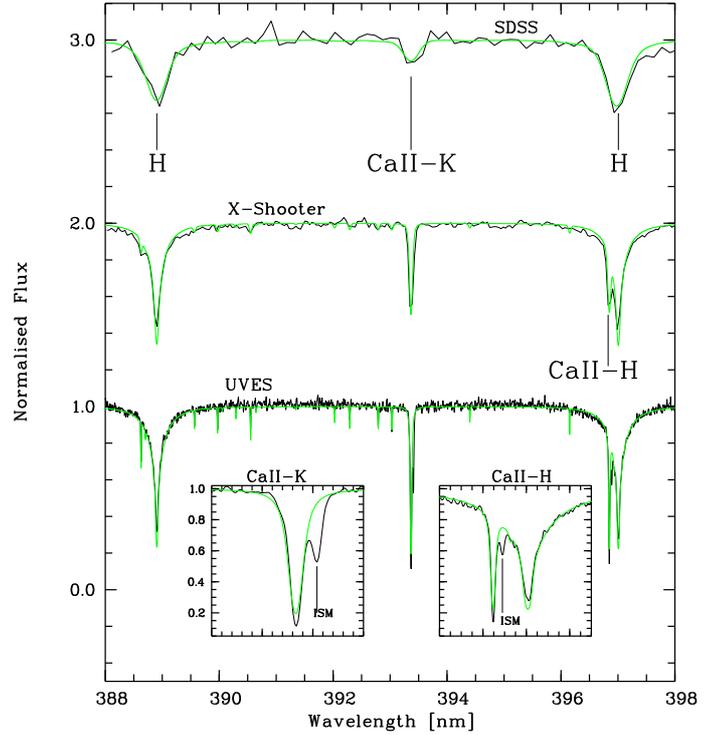}}
\end{center}
\caption[]{Range of the \ion{Ca}{ii} H and K lines.
From top to bottom, SDSS, X-Shooter, and UVES spectrum
(solid black), overimposed the synthetic profile with metallicity
-4.5, $\alpha$-enhanced by 0.4\,dex (solid green).
}
\label{caii_102915}
\end{figure}

The UVES spectrum resolves the stellar and IS components of the \ion{Ca}{ii}-K 
and \ion{Ca}{ii}-H line (see Fig.\,\ref{caii_102915}).
The EW of the stellar \ion{Ca}{ii}-K line is of 27.7\,pm, 
corresponding to an abundance of [Ca/H]=--4.47.
We did not take this line as an abundance indicator, because it is difficult
to separate the stellar and IS components,
and because with an EW of 27.7\,pm this line is saturated and
not very sensitive to abundance.
 
In the UVES spectrum we can see lines of iron peak elements 
(\ion{Fe}{i}, \ion{Ni}{i}) and
$\alpha$-elements 
(\ion{Mg}{i}, \ion{Si}{i}, \ion{Ca}{i}, \ion{Ca}{ii}, \ion{Ti}{ii}).
For the light elements Li and C-N we were unable to find an evident signature
in the spectra, therefore we can provide only an upper-limit.

For the abundance determination we relied on line profile fitting
because some lines happen to be blended (sometimes several lines
of the same element) and some lines lie on the wings of
hydrogen lines.
We computed a grid of synthetic spectra with the effective temperature and gravity
of the star, varying in [Fe/H] by 0.2\,dex.
We fitted the \ion{Fe}{i} features to derive the 1D-LTE [Fe/H].
To derive the abundances of the other elements, we computed grids of
synthetic spectra with fixed [Fe/H], by varying the abundance
[X/Fe] of the element X by 0.2\,dex and then by fitting the line profiles. 
The abundances derived for all elements are reported in Table\,\ref{102915}.
The differences with respect to Table\,1 of \citet{nature} are that we use a different 
3D model atmosphere and the applications of NLTE.

\subsection{The Li abundance}

A 3D-NLTE \citep{sbordone10} Li abundance of 2.2 (Spite plateau) would imply 
an EW for the Li doublet at 670.7\,nm of about 4.7\,pm in this star.
Such a feature should be visible in the observed spectra, but no sign of 
the line is detectable in the expected wavelength range.
In the X-Shooter spectrum, taking into account its signal-to-noise ratio 
(S/N) and the resolution, 
we expect according Cayrel's formula \citep{cayrel88}
that the limit for a feature to be visible is about 1.5\,pm ($3\sigma$),
which would correspond to a A(Li)=1.7, 
close to the Li abundance derived for the cooler
component of the binary system CS\,22876-32 \citep{gonzalez08}. 
From the S/N of the UVES spectrum (160)  an upper limit
on the EW of 0.1\,pm implies A(Li)$< 1.1$ at 
$5\sigma$ or  
A(Li)$<0.9$ at $3\sigma$.

This implies that the star is far below the Spite plateau. 
This may be linked to the fact that at extremely low metallicities the Spite
plateau displays a ``meltdown'' \citep{sbordone10}, 
i.e. an increased scatter and a lower mean 
Li abundance. This meltdown is clearly visible in the two components of the 
extremely metal-poor binary system CS\,22876-32 
([Fe/H]=--3.6, the primary with an effective temperature of 6500\,K, 
the secondary of 5900\,K), 
which show a different Li content \citep{gonzalez08}. 
The primary lies on the Spite plateau, 
while the secondary lies below at A(Li)= 1.8. 
The reasons for this meltdown are not understood. 
It has been suggested that a Li
depletion mechanism, whose efficiency is metallicity-dependent, 
might be able to explain the observations. If this
were the case, the Li abundance in SDSS\,J102915+172927 would result 
from an efficient Li depletion
caused by a combination of extremely low metallicity and 
relatively low temperature. 
If the star were a horizontal branch star \citep{hansen11}, it would be 
normal for it to be Li depleted.
However, we already argued that low gravities, 
compatible with an HB status, 
are ruled out. A sub-giant status should not imply a high Li depletion.
The absence of Li could be explained if SDSS\,J102915+172927 were a 
``blue straggler to be'' \citep{ryan}. In this
case we would expect a measurable line broadening owing to rotation. 
In our UVES spectra we cannot
derive any line broadening above what is caused by the instrumental resolution, 
which is set by the seeing. Therefore
all available evidence suggests that SDSS\,J102915+172927 is in an 
evolutionary status from the main sequence to the sub-giant branch.

\subsection{CNO upper limits}

\begin{figure}
\begin{center}
\resizebox{\hsize}{!}{\includegraphics[draft = \draftflag, clip=true]%
{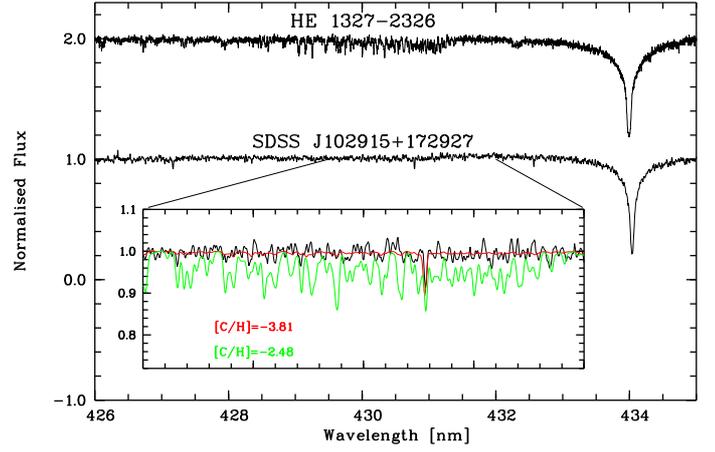}}
\end{center}
\caption[]{G-band range of the the star SDSS\,J102915+172927:
the observed spectrum (solid black) is compared to the known 
C-enhanced star HE 1327-2326.
A zoom on the G-band is shown (solid black) compared to the best
fit of the range (solid red) and with a synthetic profile
enhanced in C by 2\,dex (solid green).
}
\label{s102915gband}
\end{figure}

No strong carbon enhancement is evident in the UVES spectrum (see Fig.\,\ref{s102915gband}). 
We fitted the range of the G-band at 430\,nm, and found [C/H]$_{\rm 1D}=-3.81$.
The S/N in the range is of about 70.
The same fit of a synthetic spectrum with S/N of 50 gives [C/H]=--4.25,
C-enhancement of 0.25\,dex with respect to the input value.
We consider [C/H]$<-3.81$ as an upper limit.
If we applied 3D corrections, the abundance decreased and 
the upper limit became more strict. 
This is because the temperature structure of a 3D model
is cooler than the reference 1D model in the external layers.
In these cooler layers, the formation of molecules is favoured,
which means that with the same abundance the observed molecular bands are deeper
which in turn implies a lower abundance, see \citet{gonzalez10} and \citet{behara} for details.

\begin{figure}
\begin{center}
\resizebox{\hsize}{!}{\includegraphics[draft = \draftflag, clip=true]%
{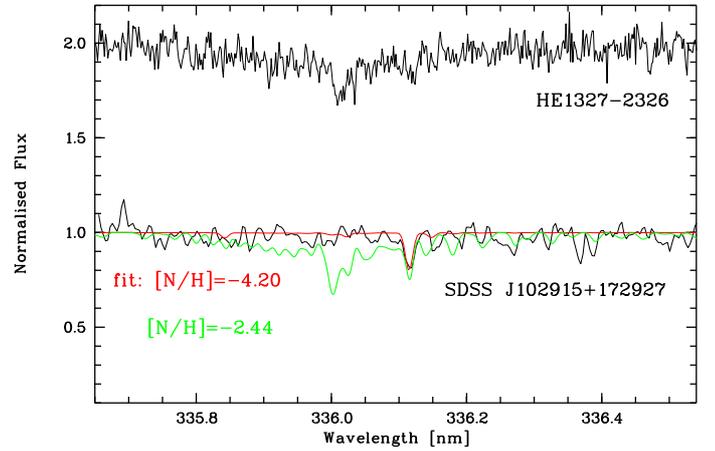}}
\end{center}
\caption[]{NH-band range of the the star SDSS\,J102915+172927:
the observed spectrum (solid black) 
is compared to the best
fit of the range (solid green) and with a synthetic profile
enhanced in N by 2\,dex (solid green).
For comparison the known 
N-enhanced star HE 1327-2326 is shown.
}
\label{s102915nhband}
\end{figure}

For the nitrogen we investigated the NH band at 336\,nm.
We found no evidence of enhancement (see Fig.\,\ref{s102915nhband}) here
either; when we fited the range we obtained an upper limit of an N-enhancement of 0.24\,dex.
As for the G-band, we consider this value as an upper limit for nitrogen.
The 3D corrections would imply a lower upper limit for this molecule as well.
 
\begin{table*}
\caption{\label{102915}
SDSS\,J102915+172927.
The abundances
[X/H] from the line profile fits are given. The adopted  \loggf\ values
 are given  in Table\ref{lines}.
}
\begin{center}
\begin{tabular}{llllllll}
\hline\noalign{\smallskip}
Element & \multicolumn{4}{c}{[X/H]$_{\rm 1D}$} & N lines & ${\rm S_{\rm H}}$ & A(X)$_\odot$ \\
       & 1D-LTE &   +3D\,cor & +NLTE & {+ 3D\,cor + NLTE } & \\
\hline\noalign{\smallskip}
 \ion{C}{ }   & $\le -3.8$      & $\le -4.5$      &                 &                 & G-band  &     & 8.50\\
 \ion{N}{ }   & $\le -4.1$      & $\le -5.0$      &                 &                 & NH-band &     & 7.86\\
 \ion{Mg}{i}  & $-4.71\pm 0.11$ & $-4.68\pm 0.11$ & $-4.52\pm 0.11$ & $-4.49\pm 0.12$ &   5     & 0.1 & 7.54\\
 \ion{Si}{i}  & $-4.27$         & $-4.30$         & $-3.93$         & $-3.96$         &   1     & 0.1 & 7.52\\
 \ion{Ca}{i}  & $-4.72$         & $-4.82$         & $-4.44$         & $-4.54$         &   1     & 0.1 & 6.33\\
 \ion{Ca}{ii} & $-4.60\pm 0.11$ & $-4.73\pm 0.03$ & $-4.82\pm 0.02$ & $-4.95\pm 0.09$ &   3     & 0.1 & 6.33\\
 \ion{Ti}{ii} & $-4.75\pm 0.18$ & $-4.83\pm 0.16$ & $-4.76\pm 0.18$ & $-4.84\pm 0.16$ &   6     & 1.0 & 4.90\\
 \ion{Fe}{i}  & $-4.73\pm 0.13$ & $-5.02\pm 0.10$ & $-4.60\pm 0.13$ & $-4.89\pm 0.10$ &  43     & 1.0 & 7.52\\
 \ion{Ni}{i}  & $-4.55\pm 0.14$ & $-4.90\pm 0.11$ &                 &                 &  10     &     & 6.23\\
 \ion{Sr}{ii} & $\le -5.10$     & $\le -5.25$     & $\le -4.94$     & $\le -5.09$     &   1     &0.01 & 2.92\\
\noalign{\smallskip}\hline\noalign{\smallskip}
\end{tabular}
\end{center}
\end{table*}

SDSS\,J102915+172927, with a metallicity of ${\rm Z}=5~10^{-5}{\rm Z}_\odot$,
is the presently known object whose composition closest resembles the
primordial one.
We computation the metallicity with the 1D
abundances derived from the UVES spectrum, the 1D C and N upper limits that we derived
by fitting the molecular bands of CH and NH, respectively, 
and assuming an enhancement in oxygen of [O/Fe]=+0.6. 

\subsection{Non-LTE effects on the element abundance determinations}\label{sec:nlte}

Our present investigation is based on the non-local thermodynamic equilibrium 
(non-LTE) line formation for six chemical species. We used the non-LTE 
methods treated in our earlier studies and documented in a number
of papers, in which atomic data and the problems of line formation
were considered in detail, i.e., in \citet[][\ion{Mg}{i}]{zhao1998_mg1}, 
\citet[][\ion{Si}{i}]{shi2008_si}, \citet[][\ion{Ca}{i}-\ion{Ca}{ii}]{mash_ca}, 
\citet[][\ion{Fe}{i}]{mash_fe}, and \citet[][\ion{Sr}{ii}]{Belyakova1997_sr2}. 
To solve the coupled radiative transfer and statistical equilibrium (SE) 
equations, we used a revised version of the
DETAIL programme \citep{detail} based on the
accelerated lambda iteration, which follows the efficient method
described by \citet{rh91,rh92}. 
The non-LTE calculations were performed for the ATLAS\,9 atmospheric structure. 

It is worth noting that we used the most 
accurate atomic data for each chemical species. 
Photoionisation cross-sections are from the 
opacity project (OP) calculations \citep{1994MNRAS.266..805S}. 
Their accuracy is estimated at the level of 10\%. This small uncertainty 
translates into an abundance error of no more than 0.01~dex. 
When no OP data are available for \ion{Sr}{ii}, we ignore photoionisation
since it affects its SE only weakly because \ion{Sr}{iii} is  
an extremely small fraction of the total  strontium atoms.
The atomic data for \ion{Ca}{ii} were updated by applying 
effective collision strengths from the $R-$matrix calculations of 
\citet{2007A&A...469.1203M}. We accounted for inelastic 
collisions both with electrons and neutral H particles in SE calculations. Hydrogen collisions 
were computed using the formula of \citet{Steenbock1984} with a scaling 
factor \kH\,= 0.1 and 1. For \ion{Ca}{i}-\ion{Ca}{ii}\, and 
\ion{Fe}{i}-\ion{Fe}{ii} our favourite is \kH\,= 0.1 as estimated empirically
 from the different influence of hydrogen atom collisions on the
different lines of a given atom in solar and stellar spectra \citep{mash_ca,mash_fe}. 
For \ion{Sr}{ii}, we applied \kH\, = 0 (no hydrogenic collisions), as 
recommended by \citet{mash_sr2}.

\begin{figure}
\resizebox{88mm}{!}{\includegraphics{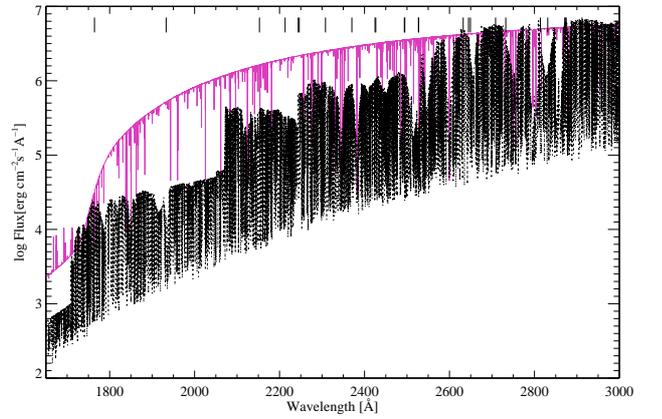}}
\caption[]{Emerging fluxes calculated using the 5811/4/$-4.5$ 
(solid black) 
and 5780/4.44/0 (dotted violet) model. Bold marks indicate the thresholds 
of the \ion{Fe}{i} levels with excitation energy of 0.9 to 3.7~eV.} \label{Fig:flux}
\end{figure}

\begin{figure}
\resizebox{88mm}{!}{\includegraphics{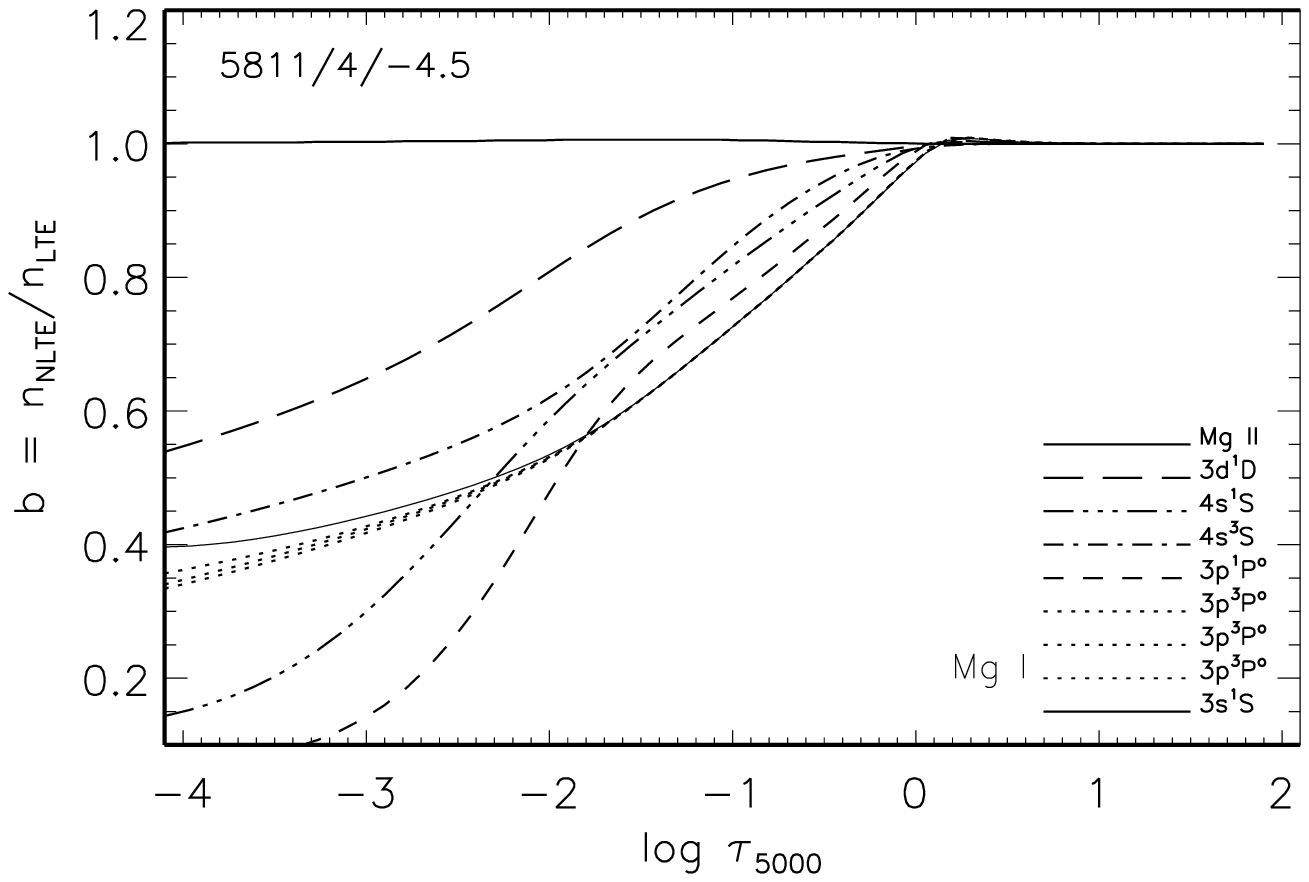}}
\vspace{-5mm}
\resizebox{88mm}{!}{\includegraphics{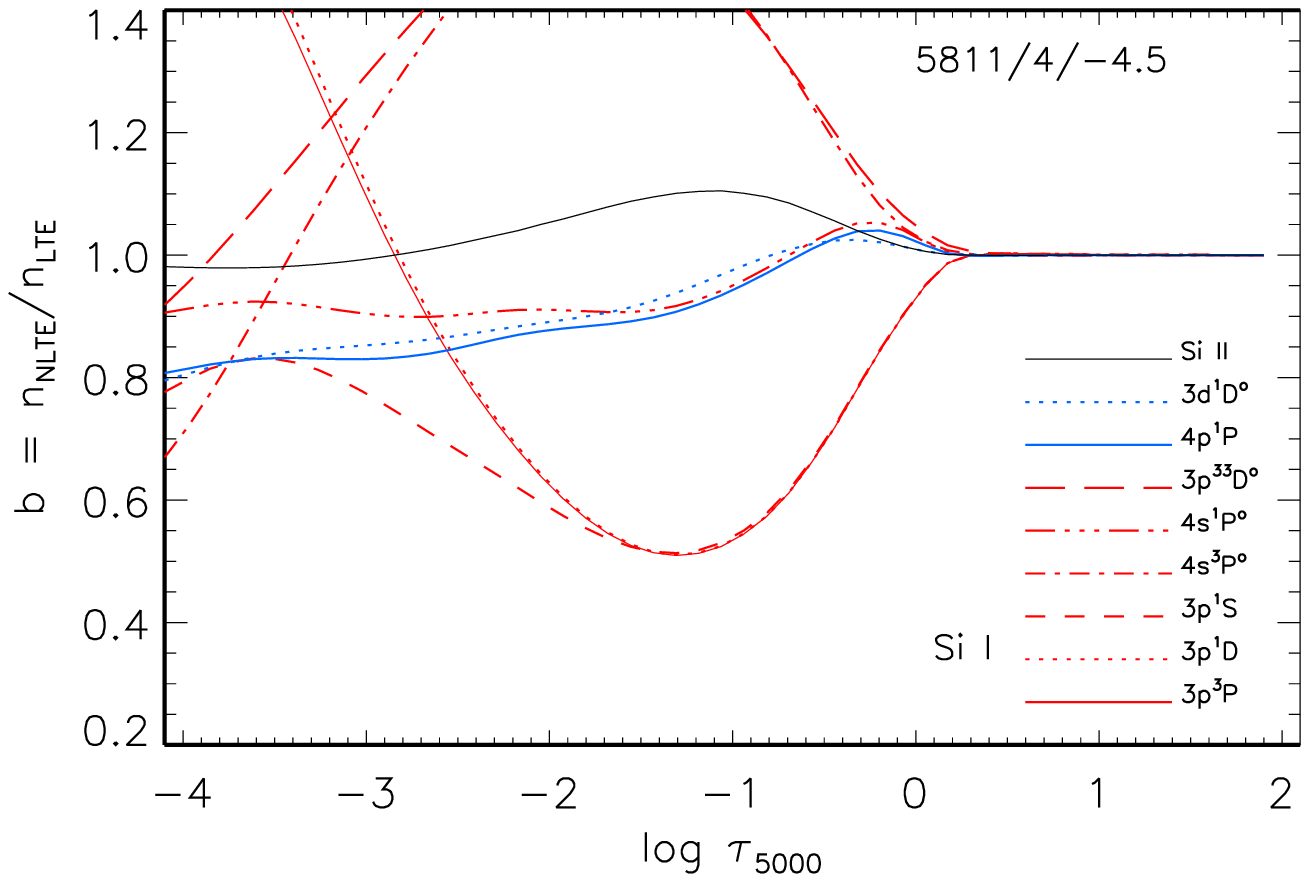}}
\vspace{-5mm}
\resizebox{88mm}{!}{\includegraphics{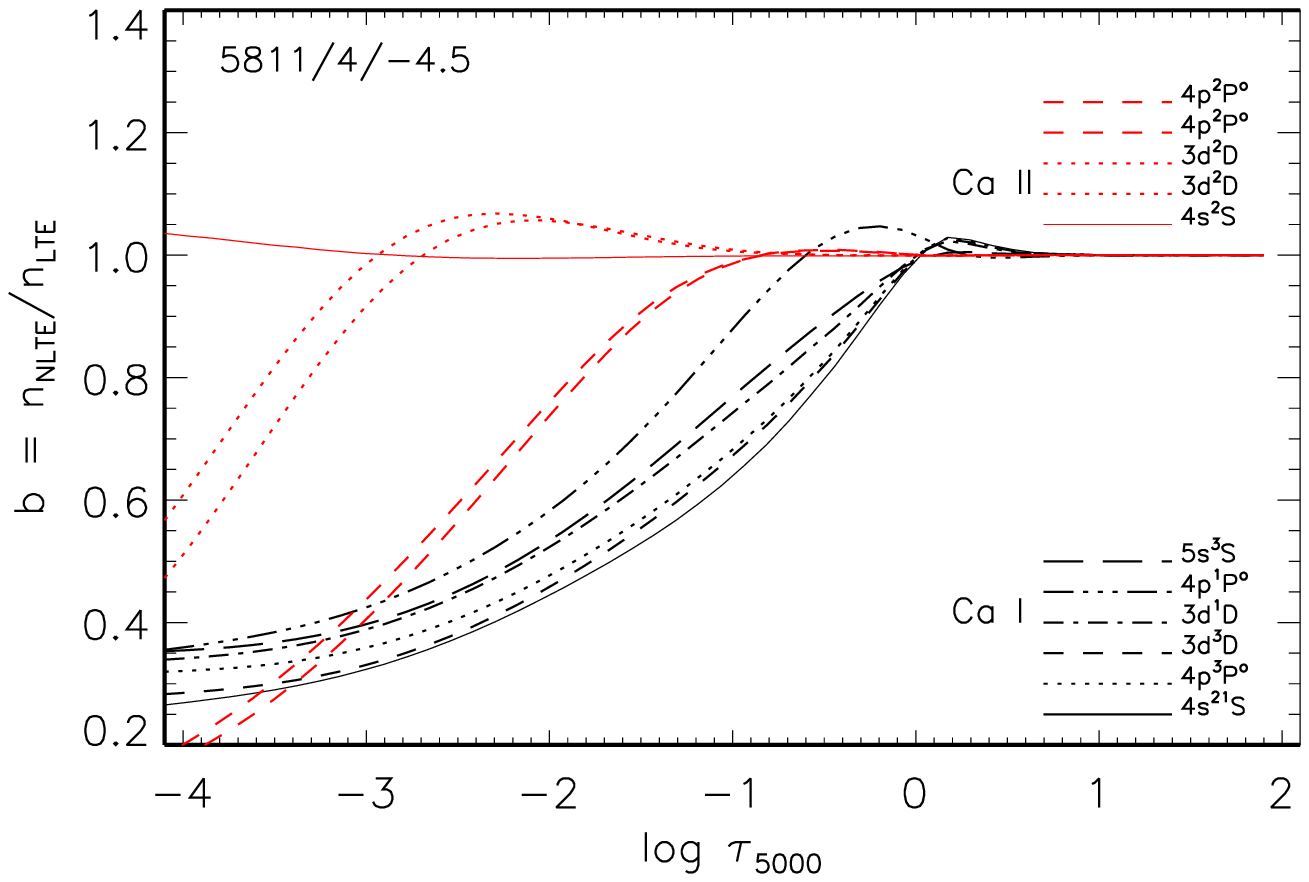}}
\vspace{-5mm}
\resizebox{88mm}{!}{\includegraphics{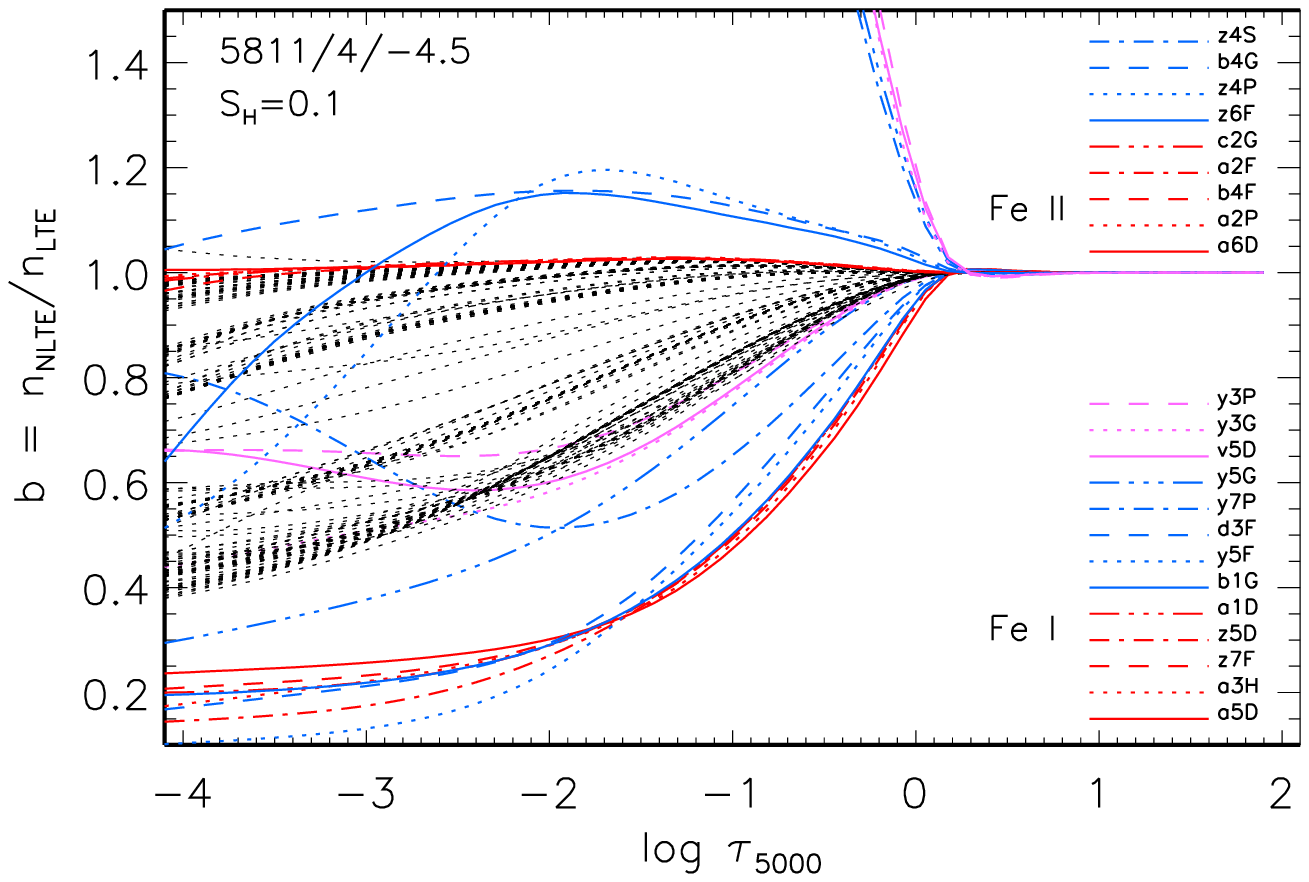}}
\\
\caption[]{Departure coefficients $b$, for the selected levels of the investigated 
atoms as a function of $\log \tau_{5000}$ in the model atmosphere 5811/4.0/$-4.5$. 
Bottom panel: we show every fifth of the first 60 levels for \ion{Fe}{i}. 
They are quoted in the bottom-right part of the panel. All remaining higher levels 
of \ion{Fe}{i} are plotted by dotted curves. We show the same for \ion{Fe}{ii}. 
They are quoted in the top-right part of the panel. 
\kH\ = 0.1 throughout.} \label{Fig:bf}
\end{figure}

The main non-LTE mechanism for the minority species in the model 5811/4.0/$-4.5$,
\ion{Mg}{i}, \ion{Si}{i}, \ion{Ca}{i}, and \ion{Fe}{i}, 
is the overionisation caused by superthermal radiation of 
non-local origin below the thresholds of the levels with \Eexc\,= 2.2-4.5~eV 
($\lambda_{thr}$ = 2240-3450\,\AA). In the extremely metal-poor atmosphere, 
deviations of the mean intensity of ionizing ultraviolet radiation from the 
Planck function are more pronounced compared with those of the solar metallicity
model (Fig.\,\ref{Fig:flux}), resulting in much stronger departures from LTE.
Figure\,\ref{Fig:bf} shows that all levels of \ion{Mg}{i}, \ion{Ca}{i}, 
and \ion{Fe}{i} and the three lowest levels of \ion{Si}{i} are strongly 
underpopulated in the line formation layers of the 5811/4.0/$-4.5$ model.  
Here we use the departure coefficients $b_i = n_i^{\rm NLTE}/n_i^{\rm LTE}$, 
where $n_i^{\rm NLTE}$ and $n_i^{\rm LTE}$ are the statistical equilibrium 
and thermal (Saha-Boltzmann) number densities, respectively. Non-LTE leads to 
a weakening of the \ion{Mg}{i}, \ion{Si}{i}, \ion{Ca}{i}, and \ion{Fe}{i} lines 
and positive non-LTE abundance corrections $\Delta_{\rm NLTE}  = \eps{NLTE}-\eps{LTE}$. 
We comment on the obtained results for individual species.

The observed \ion{Mg}{i} lines arise in the transitions 
\eu{3p}{3}{P}{\circ}{} - \eu{3d}{3}{D}{}{} (382.9-383.8~nm) and 
\eu{3p}{3}{P}{\circ}{} - \eu{4s}{3}{S}{}{} (517.2, 518.3~nm). 
For each line, the upper level is depleted to a lesser extent with regard 
to its LTE population than the lower level. Therefore, the line is weaker 
compared with its LTE strength not only because of the general overionisation 
($b_l < 1$), but also because the line source function ($S_{lu} 
\simeq b_u/b_l\,B_\nu$) rises above the Planck function ($B_\nu$) in the line 
formation layers. Here, $b_u$ and $b_l$ are the departure coefficients of 
the upper and lower levels, respectively. All investigated lines have 
similar non-LTE abundance correction at the level of +0.2~dex from the 
calculations with \kH\,= 0.1 ( Table\,\ref{Tab:dnlte}). As expected, the 
departures from LTE are reduced for increased \ion{H}{i} collision rates (\kH\,= 1). 

The effect of $b_u/b_l > 1$ resulting in $S_{lu} > B_\nu$ is more prominent 
for the only available line of silicon, \ion{Si}{i} 390.5~nm. Its lower level 
\eu{3p}{1}{S}{}{} follows the ground state of \ion{Si}{i} inside 
$\log\tau_{5000} < -1.5$ due to collisional coupling, and it is strongly 
underpopulated in the line formation layers. Its coupling to the high-excitation levels
for the upper level \eu{4s}{1}{P}{\circ}{} turns out to be
stronger than a coupling to the lower excitation levels, and tends 
to follow the continuum, \ion{Si}{ii}. This explains why \ion{Si}{i} 390.5~nm 
has a larger non-LTE correction of $\Delta_{\rm NLTE}$ = 0.34~dex (\kH\,= 0.1) 
compared to the corresponding values for the \ion{Mg}{i} lines and why 
$\Delta_{\rm NLTE}$ is only slightly reduced when moving to \kH\,= 1 (Table\,\ref{Tab:dnlte}). 

For the resonance line of \ion{Ca}{i} at 422.6~nm, the non-LTE mechanisms 
are very similar to those of the \ion{Mg}{i} lines. Calcium is the only element 
observed in SDSS\,J102915+172927 in two ionisation stages. \ion{Ca}{ii} dominates the element number
density over atmospheric depths. Therefore, no process seems to affect
the \ion{Ca}{ii} ground-state population, and $4s$ keeps its thermodynamic 
equilibrium value. The levels $3d$ and $4p$ follow the ground state in deep layers, and their
coupling is lost at the depths outside $\log\tau_{5000} < -1$ where photon
losses in the weakest line 849.8~nm of the multiplet $3d - 4p$ start to 
become important. In these atmospheric layers, $b_u/b_l < 1$ is valid for 
each investigated line of \ion{Ca}{ii} resulting in dropping the line source 
function above the Planck function and enhanced line absorption. 
For the resonance line \ion{Ca}{ii} 393.3~nm, departures from LTE occur only 
in the very core and $\Delta_{\rm NLTE}$ amounts to $-0.07$~dex. 
Non-LTE correction is larger in absolute value for the IR lines of multiplet 
$3d - 4p$, 849.8, 854.2, and 866.2 because of the overpopulation of the lower level.

\begin{figure}
\resizebox{88mm}{!}{\includegraphics{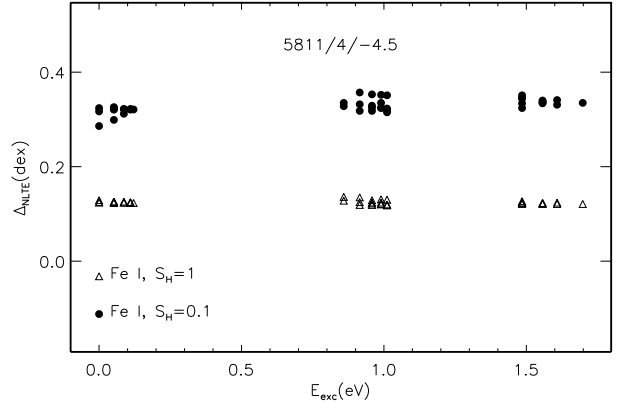}}
\caption[]{Non-LTE abundance corrections for the \ion{Fe}{i} lines 
in the 5811/4.0/-4.5 model depending on \Eexc\ from the calculations with 
\kH\ = 0.1 (filled circles) and \kH\ = 1 (triangles).} \label{Fig:dnlte_fe}
\end{figure}

Weakening of the \ion{Fe}{i} lines is mainly due 
to overionisation. In SDSS\,J102915+172927, we measured only the 
low-excitation \ion{Fe}{i} lines, with \Eexc\,= 0-1.5\,eV.  
The source function is quite similar to the Planck function for each 
investigated line, because all levels with \Eexc\,= 0-4.5~eV behave 
similarly (Fig.\,\ref{Fig:bf}). With the very similar behaviour of the departure 
coefficients for the lower levels, we calculated very similar non-LTE corrections, 
as can be seen in Fig.\,\ref{Fig:dnlte_fe}. $\Delta_{\rm NLTE}$ varies between 
0.29 and 0.36~dex in the calculations \kH\,= 0.1. Similarly to the \ion{Mg}{i} lines, 
the departures from LTE are reduced significantly for \kH\,= 1,
see Table\,\ref{Tab:dnlte}. 

Although only an upper limit was estimated for the Sr abundance, 
we performed the non-LTE calculations for \ion{Sr}{ii} with [Sr/Fe] = $-5.1$. 
Non-LTE leads to a weakened \ion{Sr}{ii} 407.7~nm line, and $\Delta_{\rm NLTE}$ 
amounts to 0.16~dex if pure electronic collisions are taken into account in the
SE calculations and decreases down to 0.12~dex for \kH\,= 1. 
For \ion{Ti}{ii}, we estimated a non-LTE correction of --0.01\,dex, 
assuming that the departures from LTE for the investigated \ion{Ti}{ii} lines 
are similar to those of the \ion{Fe}{ii} lines of similar excitation energy and equivalent width.

\begin{table*} 
 \caption{\label{Tab:dnlte} Non-LTE abundance corrections (dex) for the 
lines in SDSS\,J102915+172927 depending on surface gravity. 
}
\begin{center}
 \begin{tabular}{lclcrrrrrr}
   \hline \noalign{\smallskip}
Species & $\lambda$& \Eexc & $W_{obs}$ & \multicolumn{2}{c}{5811/3.5/-4.5} &
 \multicolumn{2}{c}{5811/4.0/-4.5} & \multicolumn{2}{c}{5811/4.5/-4.5} \\
        & (nm)     & (eV)  & (pm)      & \kH\ = 0.1 & \kH\ = 1 & \kH\ = 0.1 & \kH\ = 1 & \kH\ = 0.1 & \kH\ = 1 \\
\noalign{\smallskip} \hline \noalign{\smallskip}
\ion{Mg}{i} & 382.9355 & 2.709 &  0.8  & 0.23 & 0.08 & 0.17 & 0.04 & 0.12 & 0.02 \\
\ion{Mg}{i} & 383.2300 & 2.709 &  bl$^1$ & 0.25 & 0.09 & 0.19 & 0.05 & 0.13 & 0.03 \\
\ion{Mg}{i} & 383.8290 & 2.709 &  bl   & 0.28 & 0.10 & 0.21 & 0.06 & 0.14 & 0.03 \\
\ion{Mg}{i} & 517.2684 & 2.712 &  1.1  & 0.23 & 0.09 & 0.19 & 0.05 & 0.13 & 0.03 \\
\ion{Mg}{i} & 518.3604 & 2.717 &  1.6  & 0.24 & 0.09 & 0.19 & 0.05 & 0.13 & 0.03 \\
\ion{Si}{i} & 390.5523 & 1.909 &  1.8  & 0.31 & 0.30 & 0.34 & 0.30 & 0.35 & 0.27 \\
\ion{Ca}{i} & 422.6728 & 0.0   &  2.4 &  0.32 & &  0.28 & &  0.24 & \\
\ion{Ca}{ii} & 393.3663 & 0.0 & 27.7 & -0.08 & & -0.07 & & -0.07 & \\
\ion{Ca}{ii} & 849.8023 & 1.692 &  2.2 & -0.12 & & -0.11 & & -0.09 & \\
\ion{Ca}{ii} & 854.2091 & 1.670 &  8.6 & -0.36 & & -0.30 & & -0.22 & \\
\ion{Ca}{ii} & 866.2141 & 1.692 &  7.0 & -0.28 & & -0.24 & & -0.18 & \\
\ion{Fe}{i} & 385.9911 & 0.0 & 6.0 & 0.40 & 0.18 & 0.32 & 0.13 & 0.23 & 0.12 \\
\ion{Fe}{i} & 360.8859 & 1.011 & 2.8 & 0.47 & 0.19 & 0.35 & 0.13 & 0.25 & 0.13 \\
\ion{Fe}{i} & 407.1738 & 1.608 & 1.3 & 0.43 & 0.18 & 0.33 & 0.12 & 0.24 & 0.12 \\
\ion{Sr}{ii} & 407.7709 & 0.0  & none & 0.18$^2$ & 0.16 & 0.16$^2$ & 0.12 & 0.14$^2$ & 0.10 \\
\noalign{\smallskip}\hline\\
\multicolumn{10}{l}{ \ $^1$ line is blended} \\
\multicolumn{10}{l}{ \ $^2$ \kH\ = 0} \\
\end{tabular}
\end{center}
\end{table*}

\section{The ISM towards the star SDSS\,J102915+172927}

\begin{figure}
\begin{center}
\resizebox{\hsize}{!}{\includegraphics[angle=0,clip=true]{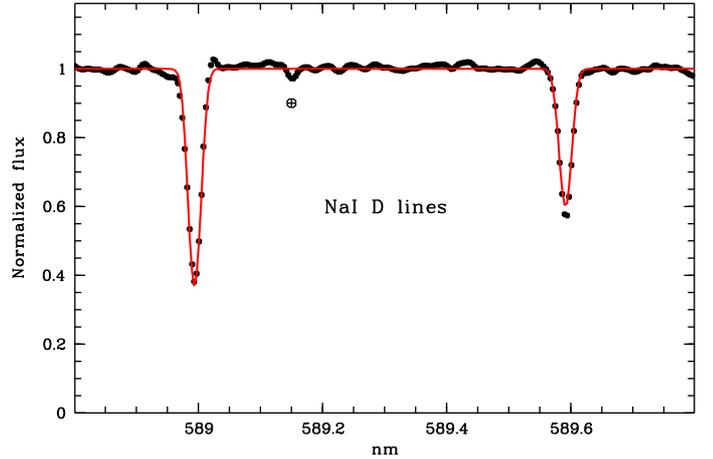}}
\end{center}
\caption[]{IS lines of \ion{Na}{i} in the spectrum of SDSS\,J102915+172927.
A telluric absorption line is denoted by a crossed circle.
The solid line is a synthesis of the IS one-component model.
}
\label{is1}
\end{figure}

The spectrum of SDSS J102915+172927 shows interstellar absorptions of the  \ion{Na}{i} D-line 
doublet at 589.0\,nm (see Fig.\,\ref{is1}) and the \ion{Ca}{ii}-K line at 393.3\,nm   
and \ion{Ca}{ii}-H line at 396.8\,nm (see detail inside Fig.\,\ref{caii_102915}).

The  interstellar feature is satisfactory modelled  with one single component 
model providing a column density of $\log$ (\ion{Na}{i}) = 12.11$\pm$0.01 cm$^{-2}$ 
and   $\log$ (\ion{Ca}{ii}) = 12.02 $\pm$ 0.04 cm$^{-2}$. 
The broadening of the lines is 7.3 $\pm$ 1.1 \kms\
in the \ion{Ca}{ii} lines and 5.2 $\pm$ 0.1 \kms in \ion{Na}{i}, suggesting 
that the turbulence is the dominant broadening factor and that the 
two ions do not sample precisely the same material with the \ion{Ca}{ii} lines 
tracing  ionised gas not detected in \ion{Na}{i}.

The \ion{Na}{i} column density is consistent with that  observed 
towards $\eta$ Leo, which at an angular distance of a few degrees shows  
$\log$ N(\ion{Na}{i})=12.08 cm$^{-2}$. 

High spectral resolution absorption observations of the
local neutral and partially ionised interstellar gas
using both the \ion{Na}{i} D-line doublet and the \ion{Ca}{ii}-K line  
allowed us to construct a picture of the distribution
of neutral gas in the local interstellar medium.
The Sun is  placed  within a low density interstellar
cavity  connected
by interstellar tunnels to other surrounding cavities.

\citet{welsh10} reported on the results of an extended survey of 
\ion{Na}{i} and \ion{Ca}{ii} absorption lines recorded
at high spectral resolution towards thousands of early-type stars mostly located
within 800 pc of the Sun.    
In particular, maps of the distribution of \ion{Na}{i} absorption
revealed that the local cavity has a 50 pc diameter and a 200 pc long extension  
in the direction of the star $\beta$ CMa and that there is an  extension of   
rarefied gas into the lower galactic halo that forms an open-ended
chimney feature.

The line of sight  towards  SDSS\,J102915+172927  has the galactic coordinates 
$l$=222.7 and $b$=56 i.e. lying in the third quadrant and at fairly high   
latitudes, intercepting low-density material. 

The observed column density ratio in the line of sight towards SDSS\,J102915+172927 is 
N(\ion{Na}{i})/N(\ion{Ca}{ii})=1.1. 
This ratio is a  diagnostic of the physical
conditions in the diffuse interstellar gas. In cold (T$\sim$30\,K) 
and dense gas clouds  most of the gas-phase Ca is
depleted onto grains, and the \ion{Na}{i}/\ion{Ca}{ii} ratio is $>>$ 1. 
In   warmer (T$\sim$1000\,K) and lower density ISM  much 
of the Ca remains in the gas phase,  and ratios of $<$ 1.0
are commonly found. For instance, for sight-line distances inside the 
local cloud  the \ion{Na}{i}/\ion{Ca}{ii} column density ratio for the warm 
(T$\sim$7000\,K) local 
interstellar gas clouds is $\sim$ 0.2 . 
The \ion{Na}{i}/\ion{Ca}{ii} ratio is thought to be  also connected to the  gas cloud velocity, such
that at velocities $\ge$ 30 km s$^{-1}$ interstellar dust grains may be 
destroyed by shocks and Ca is
liberated into the gas phase \citep{routly52}. 
Accordingly, a low value of the \ion{Na}{i}/\ion{Ca}{ii} ratio could be due
either to the presence of warm and partially ionised gas and/or the presence 
of interstellar shocks. Since we do not observe high velocities, the observed 
low \ion{Na}{i}/\ion{Ca}{ii} ratio is most probably caused by
significant amounts of partially ionised \ion{Ca}{ii} gas.   

The \ion{Na}{i}/\ion{Ca}{ii} 
ratio value generally falls in the range 0.5 to 20  but with  
a smaller range of ratio values, between  0.1 to 5, in quadrant 3
where our line of sight lies, and it matches  the value measured towards  
SDSS\,J102915+172927.

By using \ion{Na}{i} as a tracer of neutral hydrogen, we can infer a total 
column density   of $\log$  N(H) = 20.38 \cm \citep{ferlet85}. 
This column density is similar to that of  $\log$  N(H) = 20.43 \cm  
directly measured in $\rho$ Leo from the Ly$\alpha$ absorption line 
by means of IUE spectra \citep{shull85}. Because this is located at a distance of    
950 pc, the gas column densities  place a  robust lower  limit to the distance of  
SDSS\,J102915+172927.

With the given hydrogen column density the extinction  is A$_v$ =0.13, 
and assuming an extinction-to-reddening ratio of   A$_v$/E(B-V) =  3.1,  
we obtain   E(B-V)=  0.04, which is quite consistent with the extinction 
value as deduced from extinction maps.

\section{Discussion}

The sample of EMP candidates selected from the SDSS archive
(see details in \citealt{caffau11}),
and observed with the high-efficiency spectrographs X-Shooter \citep{caffau11},
or UVES \citep{uves-sdss}
is confirmed by this discovery to contain stars
at the lowest levels of metallicity observed in the Galaxy.
The few stars selected from SDSS and observed at high resolution
cannot allow us to assess the real efficiency of the method,
but this could be achieved with more observations.

This discovery gives our understanding of the formation of the first 
generations of stars a new perspective. The level of chemical enrichment 
of a star is characterised by the global metallicity Z. The
most metal-poor stars formerly known had a metallicity of about
${\rm Z}=10^{-3.5}{\rm Z}_\odot$, while SDSS\,J102915+172927
has a metallicity ${\rm Z}\le 10^{-4.0}{\rm Z}_\odot$.
A few stars with extremely low iron abundances (of the order of [Fe/H]=--5.0
or lower, \citealt{christlieb02,frebel05,norris07}) 
were discovered in the past decade. 
However, their chemical composition is highly
peculiar, with C, N, and O being more abundant than expected with respect to iron 
by several orders of magnitude,
corresponding to a moderately low Z, of the 
order of $10^{-2}{\rm Z}_\odot$.
This state of affairs supported the
notion that the first generations of stars were formed exclusively 
by extremely massive stars (now long extinct) and that there exists a 
critical metallicity below which low-mass stars cannot form 
\citep{bromm03,schneider03}. The
existence of SDSS\,J102915+172927 demonstrates that either such a critical 
metallicity does not exist, in line with the scenario
of \citet{greif11},
or that it is lower than $10^{-4}{\rm Z}_\odot$, in line
with the scenario of \citet{schneider03}. 
The peculiar chemical composition of HE\,0107-5240
and HE\,1327-2326
is consistent with the picture by \citet{frebel07}, using
the \citet{bromm03} theory, 
in which the critical parameter 
is not metallicity, but a suitable
combination of C and O abundances. The fact that SDSS\,J102915+172927 is 
not carbon-enhanced places this star in what this theory calls the 
``forbidden zone''. We have no upper limit on the oxygen
abundance, but because C, N, and Mg are not enhanced 
(as they are in HE\,0107-5240 and HE 1327-2326), there is no reason to suspect 
any significant O enhancement. If we assume, conservatively,
$\left[{\rm O/H}\right]\le -4.1$ and $\left[{\rm C/H}\right]\le -3.8$,
the transition discriminant\footnote{$D_{trans} = \log (10^{\rm [C/H]}+0.3\times 10^{\rm [O/H]})$} is ${\rm D}\le -3.7$,
while low-mass star formation should only be possible for ${\rm D}\ge -3.5$.
Our discovery will a give new impulse
to the formation theories of low-metallicity
stars. The notion of critical metallicity  or metallicity discriminant 
may have to be  revisited.

The properties of SDSS\,J102915+172927 are compatible with the
protogalaxy scenario discussed
in \citet{omukai08}. In this case, metal-poor gas under extreme UV irradiation,
with a normal dust-to-gas ratio and with a metallicity 
of Z$>5\times 10^{-6}$ of Z$_\odot$, 
fragments into a dense cluster of sub-solar mass stars (see their figure\, 5).  
This particular way of forming
low-mass metal-poor stars does not require any C/N/O enhancement, because of 
the peculiar thermal evolution of the gas.
Also \citet{schneider11} expect formation of stars as metal-poor as
SDSS\,J102915+172927, and the cooling of the gas cloud is due to dust.
In both cases a minimum metallicity is expected for low-mass 
star formation. 
On the other hand, the fragmentation found by the simulations
of \citet{clark11} relies on H$_2$ cooling and requires no
metal enrichment. If the IMF of the primordial stars is indeed
flat, as found by \citet{greif11}, then we may expect to find
low-mass stars even more metal-poor than SDSS\,J102915+172927 and,
possibly, of primordial metallicity. 
One concern on the observability of these primordial
stars is the possibility that their atmospheres may have been 
polluted by metals during encounters with molecular
clouds during their long lifetimes. 
While early estimations of this process predicted
significant pollution to take place
\citep{yoshii}, subsequent observations
failed to detect  the predicted ``accretion plateau''
of light elements \citep{molaro07,boesgaard,smiljanic,duncan,glopez}.
More recent estimations of the accretion
process efficiency \citep{frebel09} seem to exclude any significant
pollution, especially if the primordial stars
have a weak solar-like wind \citep{jk11}. 
These considerations should encourage an extensive search
for more stars that are as metal-poor as  SDSS\,J102915+172927
or even more metal-poor.

Increasing the number of stars at these very low metallicities
will also help understanding the behaviour of lithium at extremely low
metallicities.
If below a given metallicity low-mass stars can only be formed by
fragmentation of larger collapsing clouds \citep{clark11,greif11}, 
a distinct possibility
is that this process results in turbulence and/or
rotation fast enough to destroy the primordial lithium originally present in the gas 
effectively during the pre-main sequence phase.
In this a scenario we expect all stars below a given
metallicity to show no lithium. As star formation begins
to proceed in a more conventional way, through metal-line cooling, 
the stars do not destroy the lithium, and the Spite plateau appears.
The Spite plateau meltdown \citep{sbordone10} could simply
mark a transition region between the two regimes of star-formation.



\begin{acknowledgements}
The authors would like to thank A. Chieffi and M. Limongi, who kindly
provided a set of  unpublished isochrones.
The authors thank M. Barbieri for providing the Galactic model for 
the orbit calculations and for many useful discussions. We also thank
P. Ochner for the observations at the Asiago telescope.
E. Caffau and P. Bonifacio wish to thank ESO for the hospitality
at ESO-Santiago provided during the preparation of the paper.
P. Bonifacio, P. Fran\c cois, M. Spite, F. Spite, R. Cayrel, B. Plez and V. Hill
acknowledge support from the Programme National
de Physique Stellaire (PNPS) and the Programme National
de Cosmologie et Galaxies (PNCG) of the Institut National de Sciences
de l'Universe of CNRS.
This paper is also based on observations collected at the Asiago Observatory (Italy).
L.Mashonkina is supported by the Presidium RAS Programme ``Origin, structure, and
evolution of cosmic objects'' (No.~P-19) and the Swiss National 
Science Foundation (SCOPES project No.~IZ73Z0-128180/1).
HGL acknowledges financial support by the Sonderforschungsbereich SFB\,881
``The Milky Way System'' (subproject A4) of the German Research Foundation (DFG).
The authors would like to thank the anonymous referee for the useful suggestions.
\end{acknowledgements}

\bibliographystyle{aa}


%
%
%
\appendix
\section{Variability (absence of)}

To verify possible magnitude variations of SDSS\,J102915+172927 either on 
a long or a short time scale we obtained a sample of images at the Asiago 
Schmidt Telescope and also checked if other images in public databases were 
available. In the last case we only found four images in the Isaac Newton Telescope 
(INT) database as part of the Wide Field Survey (WFSur) of the Northern Hemisphere. 

\subsection{Asiago Schmidt Telescope}
On the night of 5 May 2011 we obtained a sequence of $4\times300$sec 
images in the V and $4\times300$sec images in the Cousins-R band 
at the 92/67 Schmidt Telescope at Cima Ekar Observatory in Asiago equipped 
with a KAI-11000M $4.0\times2.6$ million pixels CCD. With a pixel size 
of $0.85"$ the camera has a field of view of $56\times30$ arcmin$^2$. 
The seeing was not particularly good, $\simeq1.9$ arcsec, but it allowed us to obtain 
data with a good sampling of the Point Spread Function. The 
data also served as a second epoch for the proper motion measurements (see below).
We corrected the raw images in the usual way with appropriate master bias and 
flat-fields. We then used DAOPHOT \citep{stetson} for extracting PSF photometry from the images.
The calibration of the data was performed in a relative way matching all 
the sources in the field with the SDSS photometry of the star. The  
R band used here is not too far from the SDSS $r$ filter, therefore we calibrated it 
against the $r,(r-i)$ SDSS magnitude and colour. We obtained a mild colour term 
of $-0.09\pm0.01$. To calibrate the V band, which is not available 
from SDSS, we used a linear combination of the SDSS $g$ and $r$ band 
appropriate for the transformation to the Johnson-Cousin system for main 
sequence stars, which can be found on the SDSS web 
page\footnote{http://www.sdss.org/DR7/algorithms/sdssUBVRITransform.html}: 
the magnitude calculated in this way is V$=16.688\pm0.005$.

The results are reported in Table\,\ref{tab:asiago} and shown in Fig.\,\ref{fig:asiago} 
where in the top panel the four V band measurements and in the bottom the four
R band magnitudes are shown. As a reference the SDSS magnitude of SDSS\,J102915+172927 is shown 
as reported in Table~\ref{star}. The average errors of the Cima Ekar Schmidt measurements are 
$\simeq0.035$ mag for the R band and $\simeq0.032$ mag for the V band. The 
systematic difference from the SDSS photometry is r$=+0.006$ mag and for V$=+0.004$. 

\subsection{INT La Palma}
\begin{figure}
\begin{center}
\resizebox{\hsize}{!}{\includegraphics[angle=270,draft = \draftflag,clip=true]
{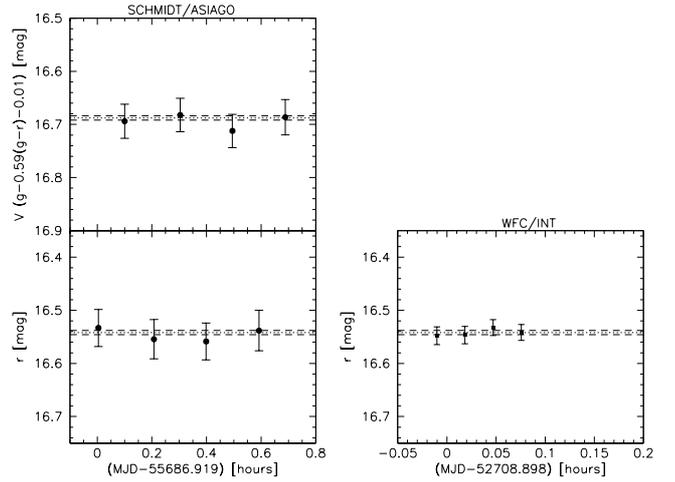}}
\end{center}
\caption[]{Magnitude measurements for SDSS\,J102915+172927 from Asiago and 
  La Palma.
  On the left column are shown the Asiago Schmidt telescope measurements versus time 
  (in hours) for the night of 5 May 2011. Top panel, the 4 V magnitude compared to a
  linear combination of the SDSS $g$ and $r$ magnitudes. Bottom panel,
  the 4 R magnitude measurements calibrated vs the SDSS $r$ band. In both
  panels the respective SDSS magnitudes are shown as short dashed lines and 
  $1\sigma$ errors (long-dashed lines). 
  On the right column are shown the $r$ band measurements of the WFC/INT on La 
  Palma for the night of 10 March 2003. }
\label{fig:asiago}
\end{figure}

Only four images in the $r$ band are present in the INT
archive of images for the Wide Field camera (WFC). These images are part of the 
WFSur\footnote{http://www.ast.cam.ac.uk/~wfcsur/index.php}, a collection of several 
programmes aiming at providing a quick coverage of the northern sky either in single- or 
multi-band. The PI of the observing programme is N. Walton, and to our knowledge the 
final catalogue and images have not yet been released but only intermediate products.
The area surrounding our star was imaged on the night of 10 March 2003 with four 
images in the $r$ band of 1 minute each. We bias- and flat-field corrected the 
images using material from the same night and produced a DAOPHOT photometry 
of only the CCD where the star is present  (CCD nr. 3). We then calibrated the raw 
WFC photometry using the 
SDSS photometry of all stars in the field, properly excluding saturated and faint 
stars. The results are shown in the right column of  Fig.\,\ref{fig:asiago} in which we 
show the variation of the magnitude in time. The plot shows the reference SDSS 
$r$-band magnitude as listed in Tab.\ref{star}. The scatter of the measurements is
$\simeq0.007$~mag with a $0.000$~mag difference with the SDSS value. 

With the material at our disposal we can safely conclude that there is 
an absence of long-term photometric variations at the level of $\Delta r\le0.006$
variations considering the errors of the star. For the short term the variations seems
to be insignificant. A longer-time monitoring with a better timing is underway to 
confirm this overall trend, considering that the monitoring covered less than 1 hour 
of data-points for the Asiago Schmidt and 12 minutes for INT data.

\begin{table}
\caption{\label{tab:asiago}
r and V photometry from Asiago Schmidt and from INT for 
SDSS\,J102915+172927.}
\begin{center}
\begin{tabular}{lcc}
\hline\noalign{\smallskip}
\multicolumn{3}{c}{Asiago/Schmidt}\\[4pt]
MJD                   & Band & Magnitude           \\[4pt]
\hline\noalign{\smallskip}
55686.919175 & r &  $16.523\pm 0.035$\\ 
55686.927659 & r &  $16.544\pm 0.037$\\ 
55686.935613 & r &  $16.549\pm 0.035$\\ 
55686.943686 & r &  $16.528\pm 0.038$\\ 
55686.923177 & V &  $16.684\pm 0.032$\\ 
55686.931663 & V &  $16.672\pm 0.032$\\ 
55686.939619 & V &  $16.702\pm 0.031$\\ 
55686.947691 & V &  $16.676\pm 0.033$\\[4pt]
\hline\noalign{\smallskip}
\multicolumn{3}{c}{WFC/INT}\\[4pt]
MJD                   & Band & Magnitude           \\[4pt]
\hline\noalign{\smallskip}
52708.897580 & r &  $16.548\pm 0.003$\\ 
52708.898767 & r &  $16.547\pm 0.002$\\ 
52708.899959 & r &  $16.533\pm 0.002$\\
52708.901151 & r &  $16.542\pm 0.002$\\[4pt]
\hline\noalign{\smallskip}
\end{tabular}
\end{center}
\end{table}

\section{Atomic data line-by-line}

{\small
\begin{table*}
\caption{\label{lines}
Lines analysed in this work.}
\begin{tabular}{llllll}
\hline\noalign{\smallskip}
Element & $\lambda$ & ${\rm E}_{\rm low}$ & \loggf & [X/H]$_{1D-LTE}$ & Notes\\
\hline\noalign{\smallskip}
 \ion{Mg}{i}  &  382.9355 & 2.709       &   -0.207               &  -4.869 & \\
 \ion{Mg}{i}  &  383.2    & 4.346/2.712 &   -0.456/+0.021        &  -4.574 &  blend\\
 \ion{Mg}{i}  &  383.8290 & 2.717       &   -1.583/+0.339/-0.408 &  -4.760 &  blend\\
 \ion{Mg}{i}  &  517.2684 & 2.712       &   -0.402               &  -4.669 & \\
 \ion{Mg}{i}  &  518.3604 & 2.717       &   -0.158               &  -4.699 & \\
 \ion{Si}{i}  &  390.5523 & 1.909       &   -1.090               &  -4.270 & \\
 \ion{Ca}{i}  &  422.6728 & 0.000       &   +0.243               &  -4.718 & \\
 \ion{Ca}{ii} &  849.8023 & 1.692       &   -1.416               &  -4.719 & \\
 \ion{Ca}{ii} &  854.2091 & 1.670       &   -0.463               &  -4.500 & \\
 \ion{Ca}{ii} &  866.2141 & 1.692       &   -0.723               &  -4.593 & \\
 \ion{Ti}{ii} &  334.9408 & 0.049       &   +0.430               &  -4.819 & \\
 \ion{Ti}{ii} &  336.1218 & 0.028       &   +0.280               &  -4.795 & \\
 \ion{Ti}{ii} &  337.2800 & 0.012       &   +0.180               &  -4.432 & \\
 \ion{Ti}{ii} &  338.3768 & 0.000       &   +0.142               &  -4.723 & \\
 \ion{Ti}{ii} &  375.9296 & 0.607       &   +0.270               &  -4.791 & \\
 \ion{Ti}{ii} &  376.1323 & 0.574       &   +0.170               &  -4.964 & \\
 \ion{Ni}{i}  &  339.2983 & 0.025       &   -0.540               &  -4.363 & \\
 \ion{Ni}{i}  &  341.4760 & 0.025       &   -0.060               &  -4.422 & \\
 \ion{Ni}{i}  &  344.6255 & 0.109       &   -0.410               &  -4.558 & \\
 \ion{Ni}{i}  &  345.8456 & 0.212       &   -0.260               &  -4.662 & \\
 \ion{Ni}{i}  &  346.1649 & 0.025       &   -0.360               &  -4.367 & \\
 \ion{Ni}{i}  &  349.2954 & 0.109       &   -0.270               &  -4.671 & \\
 \ion{Ni}{i}  &  351.0332 & 0.212       &   -0.670               &  -4.627 & \\
 \ion{Ni}{i}  &  351.5049 & 0.109       &   -0.260               &  -4.670 & \\
 \ion{Ni}{i}  &  352.4535 & 0.025       &   -0.030               &  -4.416 & \\
 \ion{Ni}{i}  &  361.9386 & 0.423       &   -0.040               &  -4.710 & \\
 \ion{Fe}{i}  &  355.8515 & 0.990       &   -0.629               &  -4.846 & \\
 \ion{Fe}{i}  &  356.5379 & 0.958       &   -0.190               &  -4.846 & \\
 \ion{Fe}{i}  &  357.0098 & 0.915       &   +0.153               &  -4.932 & \\
 \ion{Fe}{i}  &  358.1193 & 0.859       &   +0.406               &  -4.780 & \\
 \ion{Fe}{i}  &  360.8859 & 1.011       &   -0.100               &  -4.842 & \\
 \ion{Fe}{i}  &  361.8768 & 0.990       &   +0.000               &  -4.847 & \\
 \ion{Fe}{i}  &  370.5566 & 0.052       &   -1.334               &  -4.522 & \\
 \ion{Fe}{i}  &  370.9246 & 0.915       &   -0.646               &  -4.595 & \\
 \ion{Fe}{i}  &  371.9935 & 0.000       &   -0.431               &  -4.466 & \\
 \ion{Fe}{i}  &  372.7619 & 0.958       &   -0.631               &  -4.642 & \\
 \ion{Fe}{i}  &  373.7132 & 0.052       &   -0.574               &  -4.645 & \\
 \ion{Fe}{i}  &  374.5561 & 0.087       &   -0.771               &  -4.633 & \\
 \ion{Fe}{i}  &  374.5899 & 0.121       &   -1.335               &  -4.660 & \\
 \ion{Fe}{i}  &  374.8262 & 0.110       &   -1.016               &  -4.720 & \\
 \ion{Fe}{i}  &  375.8233 & 0.958       &   -0.030               &  -4.799 & \\
 \ion{Fe}{i}  &  376.3789 & 0.990       &   -0.240               &  -4.725 & \\
 \ion{Fe}{i}  &  376.7192 & 1.011       &   -0.390               &  -4.812 & \\
 \ion{Fe}{i}  &  378.7880 & 1.011       &   -0.860               &  -4.702 & \\
 \ion{Fe}{i}  &  381.5840 & 1.485       &   +0.240               &  -4.814 & \\
 \ion{Fe}{i}  &  382.0425 & 0.859       &   +0.120               &  -4.819 & \\
 \ion{Fe}{i}  &  382.4444 & 0.000       &   -1.360               &  -4.581 & \\
 \ion{Fe}{i}  &  382.5881 & 0.915       &   -0.040               &  -4.901 & \\
 \ion{Fe}{i}  &  382.7822 & 1.557       &   +0.060               &  -4.954 & \\
 \ion{Fe}{i}  &  383.4222 & 0.958       &   -0.302               &  -4.782 & \\
 \ion{Fe}{i}  &  384.0437 & 0.990       &   -0.506               &  -4.778 & \\
 \ion{Fe}{i}  &  384.1048 & 1.698       &   -0.050               &  -4.771 & \\
 \ion{Fe}{i}  &  384.9966 & 1.011       &   -0.970               &  -4.590 & \\
 \ion{Fe}{i}  &  385.6371 & 0.052       &   -1.290               &  -4.647 & \\
 \ion{Fe}{i}  &  385.9911 & 0.000       &   -0.710               &  -4.592 & \\
 \ion{Fe}{i}  &  388.6282 & 0.052       &   -1.080               &  -4.591 & \\
 \ion{Fe}{i}  &  389.5656 & 0.110       &   -1.670               &  -4.611 & \\
 \ion{Fe}{i}  &  389.9707 & 0.087       &   -1.530               &  -4.579 & \\
 \ion{Fe}{i}  &  392.2912 & 0.052       &   -1.650               &  -4.648 & \\
 \ion{Fe}{i}  &  392.7920 & 0.110       &   -1.590               &  -4.602 & \\
 \ion{Fe}{i}  &  393.0297 & 0.087       &   -1.590               &  -4.463 & \\
 \ion{Fe}{i}  &  404.5812 & 1.485       &   +0.280               &  -4.831 & \\
 \ion{Fe}{i}  &  406.3594 & 1.557       &   +0.070               &  -4.851 & \\
 \ion{Fe}{i}  &  407.1738 & 1.608       &   -0.020               &  -4.823 & \\
 \ion{Fe}{i}  &  427.1760 & 1.485       &   -0.160               &  -4.830 & \\
 \ion{Fe}{i}  &  430.7902 & 1.557       &   -0.070               &  -4.899 & \\
 \ion{Fe}{i}  &  432.5762 & 1.608       &   -0.010               &  -4.884 & \\
 \ion{Fe}{i}  &  438.3545 & 1.485       &   +0.200               &  -4.735 & \\
 \ion{Fe}{i}  &  440.4750 & 1.557       &   -1.400               &  -4.760 & \\
 \ion{Sr}{ii} &  407.7709 & 0.000       &   +0.167               &  -5.139 & \\
\hline\noalign{\smallskip} 
\end{tabular}
\end{table*}
}


\bibliographystyle{aa}

\end{document}